\documentclass[12pt]{article}

\usepackage{amsmath,amssymb,latexsym,mathrsfs}

\newlength{\dinwidth}
\newlength{\dinmargin}
\begin{document}

\renewcommand{\square}{\vrule height 1.5ex width 1.2ex depth -.1ex }


\newcommand{\II}{\leavevmode\hbox{\rm{\small1\kern-3.8pt\normalsize1}}}

\newcommand{\CC}{{\mathbb C}}
\newcommand{\RR}{{\mathbb R}}
\newcommand{\NN}{{\mathbb N}}
\newcommand{\QQ}{{\mathbb Q}}
\newcommand{\ZZ}{{\mathbb Z}}


\newcommand{\CoinfM}{C_0^\infty(M)}
\newcommand{\CoinfN}{C_0^\infty(N)}
\newcommand{\Coinfd}{C_0^\infty(\RR^d\backslash\{ 0\})}
\newcommand{\Coinf}[1]{C_0^\infty(\RR^{#1}\backslash\{ 0\})}
\newcommand{\CoinX}[1]{C_0^\infty({#1})}
\newcommand{\Coin}{C_0^\infty(0,\infty)}


\newtheorem{Thm}{Theorem}[section]
\newtheorem{Def}[Thm]{Definition}
\newtheorem{Lem}[Thm]{Lemma}
\newtheorem{Prop}[Thm]{Proposition}
\newtheorem{Cor}[Thm]{Corollary}

\numberwithin{equation}{section}


\newcommand{\DD}{{\mathscr D}}
\newcommand{\EE}{{\mathscr E}}
\newcommand{\HH}{{\mathscr H}}
\newcommand{\KK}{{\mathscr K}}
\newcommand{\FF}{{\mathscr F}}
\newcommand{\SSS}{{\mathscr S}}

\newcommand{\DDsp}{{\mathscr D}_{\rm sp}}
\newcommand{\DDco}{{\mathscr D}_{\rm cosp}}
\newcommand{\DDd}{{\mathscr D}_{\rm double}}
\newcommand{\OO}{{\cal O}}
\newcommand{\Aa}{{\cal A}}
\newcommand{\Ff}{{\cal F}}
\newcommand{\cI}{{\cal I}}
\newcommand{\cR}{{\cal R}}
\newcommand{\Uu}{{\cal U}}
\newcommand{\Xx}{{\cal X}}
\newcommand{\Zz}{{\cal Z}}
\newcommand{\Ft}{{\widetilde{\Ff}}}

\newcommand{\gb}{{\boldsymbol{g}}}
\newcommand{\hb}{{\boldsymbol{h}}}
\newcommand{\kb}{{\boldsymbol{k}}}
\newcommand{\nb}{{\boldsymbol{n}}}
\newcommand{\xb}{{\boldsymbol{x}}}
\newcommand{\xbo}{{\boldsymbol{x_0}}}
\newcommand{\etb}{{\boldsymbol{\eta}}}


\newcommand{\Wf}{{\mathfrak W}}
\newcommand{\Af}{{\mathfrak{A}}}

\newcommand{\Dal}{\fbox{\phantom{${\scriptstyle *}$}}}

\newcommand{\Ran}{{\rm Ran}\,}
\newcommand{\spec}{{\rm spec}\,}
\newcommand{\supp}{{\rm supp}\,}
\newcommand{\Span}{{\rm span}\,}
\newcommand{\Tr}{{\rm Tr}\,}
\renewcommand{\Re}{{\rm Re}\,}
\renewcommand{\Im}{{\rm Im}\,}

\newcommand{\ip}[2]{{\langle #1\mid #2\rangle}}
\newcommand{\ket}[1]{{\mid #1\rangle}}
\newcommand{\bra}[1]{{\langle #1 \mid}}
\newcommand{\stack}[2]{\substack{#1 \\ #2}}

\newcommand{\ub}{\overline{u}}
\newcommand{\vb}{\overline{v}}
\newcommand{\wb}{\overline{w}}
\newcommand{\Bb}{\overline{B}}
\newcommand{\Pb}{\overline{P}}
\newcommand{\Qb}{\overline{Q}}
\newcommand{\Xb}{\overline{X}}
\newcommand{\Xib}{\overline{\Xi}}
\newcommand{\Xxb}{\overline{\Xx}}
\newcommand{\Yb}{\overline{Y}}

\newcommand{\kt}{\widetilde{k}}

\newcommand{\fhat}{\widehat{f}}
\newcommand{\hhat}{\widehat{h}}

\newcommand{\WF}{{\rm WF}\,}

\newcommand{\Had}{{\rm Had}\,}
\newcommand{\Wt}{\widetilde{W}}

\newcommand{\LL}{{\mathcal L}}
\newcommand{\Lorpror}{{\mathscr L}^{\uparrow}_+} 

\newcommand{\bgamma}{\boldsymbol{\gamma}}

\newcommand{\omABV}{\omega_{2AB(V)}}
\newcommand{\omABVl}{\omega^{\rm loc}_{2AB(V)}}

\newcommand{\dirop}{{\boldsymbol{\nabla}}\!\!\!\!\!/ \,}
\newcommand{\dualdirop}{\overset{*}{{\boldsymbol{\nabla}}\!\!\!\!/}}
\newcommand{\Sco}{S_{\rm cosp}}
\newcommand{\Ssp}{S_{\rm sp}}
\newcommand{\Nn}{\mathcal{N}}

\newcommand{\TT}{\mathcal{T}}
\newcommand{\WW}{\mathscr{W}}

\begin{titlepage}
\renewcommand{\thefootnote}{\fnsymbol{footnote}}

\LARGE
\center{\bf A Quantum Weak Energy Inequality \\ for Dirac Fields in
  Curved Spacetime}
\Large

\vspace{0.2in}
\center{Christopher J. Fewster${}^1$ and Rainer Verch${}^2$}
\normalsize
\center{${}^1$ Department of Mathematics, University of York, \\
Heslington, York YO10 5DD, United Kingdom. \\[4pt]
e-mail: {\tt cjf3@york.ac.uk}\\ 
\vspace{0.1in}
$^2$ Institut f\"ur Theoretische Physik,
                 Universit\"at G\"ottingen,\\
                 Bunsenstr.\ 9, 
                 D-37073 G\"ottingen, Germany.\\[4pt]
                 e-mail: {\tt verch$@$theorie.physik.uni-goettingen.de}}
\normalsize
\center{May 21, 2001}

\begin{abstract}
Quantum fields are well known to violate the weak energy condition of
general relativity: the renormalised energy density at any given point
is unbounded from below as a function of the quantum state. By contrast,
for the scalar and electromagnetic fields it has been shown that 
weighted averages of the energy density along timelike curves satisfy
`quantum weak energy inequalities' (QWEIs) which constitute lower bounds
on these quantities. Previously, Dirac QWEIs have been obtained only for
massless fields in two-dimensional spacetimes. 
In this paper we establish QWEIs for the Dirac and Majorana fields of mass $m\ge 0$ 
on general four-dimensional globally hyperbolic spacetimes, averaging
along arbitrary smooth timelike curves with respect to any of a
large class of smooth compactly supported positive weights.
Our proof makes essential use of the microlocal characterisation of the
class of Hadamard states, for which the energy density may be defined by
point-splitting. 
\end{abstract}

\end{titlepage}

\setcounter{footnote}{0}

\section{Introduction}\label{sect:intro}

In general relativity, it is customary to assume that the stress-energy
tensor satisfies one or more of the classical energy conditions; 
the weak energy condition, for example, being the assertion that the
energy density 
measured by any observer is nonnegative. The primary motivation behind these
energy conditions is that they ensure that gravity acts as an attractive force
(in the sense of focussing geodesic congruences) in accordance with our
experience of gravitation on a wide range of scales. It is therefore 
natural to assume that physically reasonable forms of classical matter 
obey such conditions, and to regard matter theories (such as the nonminimally
coupled scalar field~\cite{Deser,BarVis,FlanWald,FordRomScalar})
violating such conditions
as being of questionable physical significance on many scales. Moreover, the
energy conditions have proved to be of great value in obtaining deep
results in classical general relativity, such as the positive mass and 
singularity theorems \cite{SY,Witten,HE}. 

However, it is well known that all the pointwise energy conditions are violated
in quantum field theory. Indeed, Epstein, Glaser and Jaffe
\cite{EGJ} proved that 
no Wightman field theory on Minkowski space can admit a (nontrivial) energy
density observable whose expectation values are bounded from below and
vanish in the Minkowski vacuum state. Moreover, in linear field theories (both
in flat and curved spacetimes) it is easy to construct states whose 
energy density at a given point may be tuned to
be arbitrarily negative~\cite{DaviesFulling,Klinkhammer}.
This raises the possibility that quantum matter 
might be used to construct spacetimes with exotic properties, such as traversable
wormholes~\cite{FR-worm} or so-called `warp drive' spacetimes~\cite{PfenFordWarp}, 
usually excluded by the classical energy conditions. One might also
ask whether the conclusions of the singularity theorems remain valid for
quantum matter. Furthermore, it is necessary to understand how classical
matter contrives to obey the classical energy conditions, given that its
fundamental constituents need not. 

One profitable line of enquiry, starting with the work of
Ford~\cite{FordNegEn},  has been to 
investigate weighted averages of the renormalised energy density along
the worldline 
of an observer, or over a small spacetime region. It turns out that the
expectation values of these averaged observables are bounded from below
independently of the state, and such bounds 
have been developed in successively greater generality over the
last few years
\cite{FordRomanRest,PfenFord,Flan2d,EvesonFewster,FewsterTeo1,Helfer,AGWQI}.
Most recently, one of us~\cite{AGWQI} has established  
the existence of such bounds (and given an explicit, though not optimal, lower
bound) for the minimally coupled real linear scalar field in any globally hyperbolic
spacetime, in the case where averaging is performed with respect to
proper time along any smooth timelike curve using an arbitrary smooth compactly supported positive
weight belonging to the class\footnote{See the remarks following
Theorem~\ref{thm:main} for a brief discussion of this class.}
\begin{equation}
\WW =\{ f\in\CoinX{\RR}\mid f(\tau)=g(\tau)^2~\hbox{for some
real-valued}~g\in\CoinX{\RR}\}\,.
\label{eq:weights}
\end{equation}

The constraints imposed by these lower bounds have been called `quantum
inequalities' by various authors. However, as we hope to discuss elsewhere,
there seem to be strong parallels between the phenomena discussed above and
various situations arising in quantum mechanics. Indeed, quantum
inequalities appear to be a widespread feature of quantum theory as a
whole, stemming ultimately from the uncertainty principle. In this
light, we adopt the more specific terminology `quantum weak energy
inequality' (or QWEI) in relation to lower bounds on the
renormalised energy density of a quantum field. 

There are also related constraints which demand that the integral of
the energy density over any complete (i.e.\ inextendible) smooth
timelike or lightlike (`null'-) curve be non-negative. These are
called the `averaged weak energy condition' (AWEC) or `averaged null energy
condition' (ANEC), respectively; they may be viewed as a limiting
case of QWEIs when the weight function ($f$ in eqn.~(\ref{eq:QIintro}) below)
against which the energy density is integrated along a timelike or
lightlike curve approaches the unit function. Historically, it was
first pointed out in a work by Tipler \cite{Tipler} that suitable
versions of AWEC or ANEC imply singularity theorems in general
relativity similar to those which one obtains from pointwise
positivity conditions on the energy density. Subsequently, the
question of whether quantum fields obey AWEC or ANEC has been investigated in
a number of works
\cite{Klinkhammer,Folacci,WaldYurt,YurtANEC1,YurtANEC2,FordRomanQInEq,FlanWald,VerchANEC}. 
Most of these references treat linear quantum fields; \cite{VerchANEC}
establishes ANEC for general (axiomatic) quantum field theory in
two-dimensional Minkowski spacetime. A study of the interrelations
between averaged energy conditions and QWEIs is contained in
\cite{FordRomanQInEq}. We refer to \cite{FlanWald} for further
discussion and review of averaged energy conditions.  

QWEIs place stringent constraints on attempts to generate exotic 
spacetimes~\cite{FR-worm,PfenFordWarp} and may open a route towards proving results
analogous to the singularity theorems for quantum matter
\cite{YurtANEC1,YurtANEC2}. To date, however, most QWEI results have been
obtained for scalar field theories, while the more physically
interesting electromagnetic and Dirac fields have received comparatively
little attention. 
Ford and Roman have considered the electromagnetic field in Minkowksi
space~\cite{FordRomanRest} and shown that a QWEI holds for averaging against
the Lorentzian weight $f(\tau)=\tau_0/[\pi(\tau^2+\tau_0^2)]$
along timelike geodesics; this result was generalised to static
trajectories in static spacetimes by Pfenning~\cite{PfThesis}, who has
also removed the restriction to Lorentzian weights~\cite{PfEM}.
It is reasonable to suppose that even more general QWEIs may be obtained
for this case. 

Both the scalar and electromagnetic fields have the
property that the classical energy density is manifestly nonnegative, a
fact which underpins all the results on these fields. The Dirac field is
technically very different in that the `classical' energy density 
is unbounded both from above and below; in second quantization,
renormalisation serves the dual purpose of restoring finiteness and
imposing positivity of the Hamiltonian. This problem appears to have
restricted progress on the Dirac field to date. The main contribution
has been that of Vollick, who established a QWEI for Dirac fields in two-dimensional 
spacetimes~\cite{Voll2} by converting the
problem to one involving a scalar field and then adapting arguments due
to Flanagan~\cite{Flan2d}. There seems little prospect of generalising
this argument beyond the two-dimensional setting. In four dimensions, 
Vollick has also given explicit examples of states with locally negative
energy densities~\cite{Voll1} and demonstrated that the resulting energy
densities nonetheless obey QWEIs modelled on those for the scalar field. 

In this paper we establish a general QWEI for massive or massless
Dirac fields on 
four-dimensional\footnote{The restriction to four dimensions is purely
for convenience: our methods would apply in more general dimensions.} 
globally hyperbolic spacetimes. To be more specific, let
$\gamma$ be a smooth timelike curve, parametrized by its proper time $\tau$,
in a globally hyperbolic spacetime
$(M,\gb)$. Let $\omega_0$ be a given (but arbitrary) Hadamard
state\footnote{See Sec.~\ref{sect:Dirac} for a brief review of the concepts used here.}
of the Dirac field on $(M,\gb)$. The state $\omega_0$ is used as a
`reference state' to define the expected normal ordered energy density
$\langle :T_{00}:\rangle_\omega$ for any other Hadamard state $\omega$. 
Our main result, Theorem~\ref{thm:main} asserts that
\begin{equation}
\inf_{\omega} \int d\tau\,\langle :T_{00}:\rangle_\omega(\gamma(\tau)) f(\tau)
>-\infty,
\label{eq:QIintro}
\end{equation}
where the infimum is taken over the class of Hadamard states and
$f$ belongs to the class $\WW$. In principle our arguments yield 
an explicit lower bound for the left-hand side
of~(\ref{eq:QIintro}). This expression is unfortunately not particularly
enlightening and is not expected to be sharp. Let us note
that~(\ref{eq:QIintro}) remains true if the normal ordered energy
density is replaced by the renormalised energy density, as these two
quantities differ by a smooth function. 
 
The plan of the paper is as follows. We begin, for completeness and to
fix notation, by reviewing the theory of the quantized Dirac field in
Sec.~\ref{sect:Dirac}. Particular attention is given to the class of
Hadamard states, which may be characterised by a microlocal spectrum
condition on the wave-front set of the two-point function. This
formulation of the Hadamard condition is technically convenient and
allows us to bring the tools of microlocal analysis to bear. 
Sec.~\ref{sect:psed} explains how the normal ordered energy density
may be constructed by point-splitting. 

The proof of our QWEI begins in 
Sec.~\ref{sect:main}, using the following strategy.
The averaged normal ordered
energy density is first expressed as an integral over $\RR^2$; decomposing this integral
according to the quadrants of $\RR^2$, each piece
is then split further into four using a decomposition induced by the
reference state $\omega_0$. All but two of the resulting sixteen
contributions can be bounded (both above and below) using estimates
obtained in Sec.~\ref{sect:pfdineq}. 
The remaining terms are then expressed
in the form $\cR=\lim_{\Lambda\to+\infty}\Tr J_\Lambda W$
where $J_\Lambda$ and $W$ are self-adjoint and $J_\Lambda$ is independent of
$\omega$. The parameter $\Lambda\in\RR^+$ defines a cut-off, used to
avoid domain problems. We prove that $W$ is positive and trace-class
with bounded trace as $\omega$ varies. To conclude that $\cR$ is bounded
below, it then suffices to establish that the operators 
$J_\Lambda$ are bounded below
uniformly in $\Lambda$. This is accomplished in Sec.~\ref{sect:J},
completing the proof of our QWEI. In Sec.~\ref{sect:Maj} we briefly
describe how our arguments can also be applied to the Majorana field. 

{\noindent\bf Conventions:} The metric signature is $(+,-,-,-)$.
Lower (resp. upper) case Latin characters from the
beginning of the 
alphabet will label tetrad (resp. spinor) indices. Tetrad indices run
from 0 to 3, and we will use $j,k$ to label the spatial components
$1,2,3$. The summation convention will be used throughout the paper except where
otherwise indicated. Units with $c=\hbar=1$ are adopted. 

The Fourier transform of an integrable function $f$ on $\RR^n$ will be
defined using the non-standard convention 
$\widehat{f}(k) = \int d^nx\, {\rm e}^{ikx}f(x)$, with 
inverse $\check{h}(x) = (2\pi)^{-n}\int d^nk\, 
{\rm e}^{-ikx}h(k)$. The Fourier transform of a distribution $u$ with
compact support is $\widehat{u}(k) = u(e_k)$ with $e_k(x) = {\rm e}^{ikx}$.

Given a Lorentzian manifold $(M,\gb)$, $\DD'(M)$ will denote the space
of distributions on $M$ as defined in \S6.3 of~\cite{Hormander1}. Thus
if $u\in\DD'(M)$ there is, for each chart $(U,\kappa)$ in $M$, a
distribution $u_\kappa\in\DD'(\kappa(U))$ such that\footnote{The factor of
$\sqrt{-|\gb|}$ is used to identify test functions with test densities,
on which $u$ strictly speaking acts.}
\begin{equation}
u(f) = u_\kappa((\sqrt{-|\gb|}f)\circ \kappa^{-1}) \qquad\forall
f\in\CoinX{U} 
\label{eq:distdef}
\end{equation}
and so that $u_\kappa=(\kappa'\circ\kappa^{-1})^*u_\kappa'$ in
$\kappa(U\cap U')$ for any other chart $(U',\kappa')$. We will also (and
usually) write $u\circ\kappa^{-1}$ for $u_\kappa$. Spinor and cospinor
distributions will be defined in an analogous fashion, with the
convention that a spinor distribution acts on test cospinor fields, and
vice versa.

\section{The Quantized Dirac Field}
\label{sect:Dirac}
\subsection{Geometrical preliminaries}
In order to make the present paper sufficiently self-contained, we
need to summarize a few basic facts about the geometry of spinor fields
in curved spacetimes. We will follow Dimock's work \cite{Dim.D} to
large extent.

We will consider Dirac fields in a four-dimensional globally hyperbolic
spacetime $(M,\gb)$. To begin with, we recall that a globally hyperbolic
spacetime is a Lorentzian spacetime admitting a Cauchy surface, the
latter being a smooth hypersurface in $M$ which is intersected exactly
once by each inextendible causal curve in $(M,\gb)$. We will also
suppose that $(M,\gb)$ is orientable and time-orientable, and that
such orientations have been chosen. Then $(M,\gb)$ possesses a
spin-structure, that is, there is a principal fibre bundle $S(M,\gb)$
having SL$(2,\CC)$ as structure group, acting from the right, together
with a 2-1 fibre-bundle homomorphism $\psi: S(M,\gb) \to F(M,\gb)$ which
projects $S(M,\gb)$ onto the frame bundle $F(M,\gb)$. That is to say,
$\psi$ preserves base-points and obeys
\begin{equation}
\psi \circ R_{\bf s} = R_{\Lambda({\bf s})} \circ \psi\,,
\end{equation}
where $R$ denotes the right action of the structure groups on the
principal fibre bundles involved and SL$(2,\CC) \owns {\bf s} \mapsto
\Lambda({\bf s}) \in \Lorpror$ is the covering projection onto the
proper orthochronous Lorentz group. We recall that $F(M,\gb)$ is the
bundle of oriented and time-oriented tetrads $(e_0,e_1,e_2,e_3)$, so
that $\gb(e_a,e_b) = \eta_{ab}$ with $\eta_{ab} = {\rm
  diag}(1,-1,-1,-1)$ where $e_0$ is timelike and future-pointing and
the tetrad is given the orientation of $M$.

Moreover, a collection of $4 \times 4$-matrices
$\gamma_0,\ldots,\gamma_3$ is called a set of Dirac matrices if
\begin{equation}
\gamma_a \gamma_b + \gamma_b \gamma_a = 2 \eta_{ab} \cdot \II\,.
\end{equation}
A theorem due to Pauli states that, if $\gamma_0,\ldots,\gamma_3$ and
$\gamma_0',\ldots,\gamma_3'$ are two sets of Dirac matrices, then
there is an invertible matrix $M$ so that $\gamma_a = M
\gamma_a'M^{-1}$.
Any set of Dirac matrices $\gamma_0,\ldots,\gamma_3$ is connected to
the covering SL$(2,\CC) \overset{\Lambda(\,.\,)}{\longrightarrow}
\Lorpror$ in the following way. Let Spin$(1,3)$ consist of all
unimodular $4 \times 4$-matrices $S$ so that
\begin{equation}
\label{GammaS}
 S\gamma_a S^{-1} = \gamma_b \Lambda^b{}_a
\end{equation} 
holds for some real numbers $\Lambda^b{}_a = \Lambda^b{}_a(S)$. It
follows from the defining properties of Dirac matrices that
$\Lambda^b{}_a(S)$ is contained in the Lorentz group. The restriction
of the map $S \mapsto \Lambda^b{}_a(S)$ in \eqref{GammaS} to ${\rm
  Spin}_0(1,3)$, the unit connected component of Spin$(1,3)$, is a
group homomorphism with range $\Lorpror$, and thus ${\rm
  Spin}_0(1,3)$ is isomorphic to SL$(2,\CC)$.

Sometimes it is useful to distinguish sets of Dirac matrices with
certain properties. One says that a set $\gamma_0,\ldots,\gamma_3$ of
Dirac matrices belongs to a standard representation if
\begin{equation}
\gamma_0^* = \gamma_0\ \ \ {\rm and} \ \ \ \gamma_k^* = -\gamma_k\,.
\end{equation}
Here, $\gamma_a^*$ is the hermitean adjoint of $\gamma_a$. A set of
Dirac matrices which belongs to a standard representation and has the
additional property that the complex conjugate matrices fulfill
\begin{equation}
 \overline{\gamma}_a = - \gamma_a\,,
\end{equation}
is said to belong to a Majorana representation.

We will now suppose that we are given a globally hyperbolic spacetime
$(M,\gb)$ together with a spin-structure $(S(M,\gb),\psi)$ and a set
of Dirac matrices $\gamma_0,\ldots,\gamma_3$ which we will assume, for
the sake of notational simplicity, to belong to a standard
representation. (We note, however, that everything which follows could
also be carried out in a similar way without that assumption.) As was
pointed out above, via the isomorphism ${\rm Spin}_0(1,3) \simeq {\rm
  SL}(2,\CC)$, $\CC^4$ carries a representation of the universal
covering group of $\Lorpror$ which is given by the action of the
matrices $S$ in ${\rm Spin}_0(1,3)$ on vectors in $\CC^4$. Via ${\rm
  Spin}_0(1,3) \simeq {\rm SL}(2,\CC)$, we can also regard $S(M,\gb)$ as
a ${\rm Spin}_0(1,3)$-principal bundle and form the associated vector
bundle
\begin{equation}
 DM = S(M,\gb) \ltimes_{{\rm Spin}_0(1,3)} \CC^4\,.
\end{equation}
That is, the fibre of $DM$ at $p \in M$ consists of the orbits
\begin{equation}
[{\bf s}_p,x] = \{(R^{-1}_S {\bf s}_p,Sx): S \in {\rm Spin}_0(1,3)\}
\end{equation}
for ${\bf s}_p \in S(M,\gb)_p$ and $x \in \CC^4$. There is a fibrewise
left action of ${\rm Spin}_0(1,3)$ on $DM$ by
$L_S[{\bf s}_p,x] = [{\bf s}_p,Sx]\,.$
Elements in $DM$ are called spinors, and elements in the dual bundle
$D^*M$ are called cospinors.
Moreover, if $E$ is a (local) section in $S(M,\gb)$, then it induces on
one hand a tetrad field $(e_0,\ldots,e_3) = \psi \circ E$, i.e.\ a
(local) smooth section in $F(M,\gb)$, via the spin-structure, and on the other
hand it induces a set $(E_A)_{A = 1}^4$ of (local) smooth sections in
$DM$, defined by
\begin{equation}
                   E_A = [E,b_A]
\end{equation}
where $b_1,\ldots,b_4$ is the standard basis in $\CC^4$. There are
corresponding dual tetrad fields $(e^0,\ldots,e^3)$ defined by
$e^b(e_a) = \delta^b_a$ and $(E^B)_{B = 1}^4$ defined by $E^B(E_A) =
\delta^B_A$.  
The $e^b$ are smooth sections in $T^*M$, and the $E^B$ are smoth
sections in $D^*M$, the dual bundle to $DM$. We shall denote by
$C^{\infty}(DM)$ and $C^{\infty}(D^*M)$ the sets of smooth sections in
$DM$ and $D^*M$, respectively. The notation for smooth sections in
$TM$ and $T^*M$ will be similar.

With respect to the given set of Dirac matrices, one can define a
section $\bgamma$ in $C^{\infty}(T^*M) \otimes C^{\infty}(DM) \otimes
C^{\infty}(D^*M)$, i.e.\ a mixed spinor-tensor field, by setting its
components $\gamma_b{}^A{}_B$ in the induced frame $e^b \otimes E_A
\otimes E^B$ to be equal to the matrix elements $(\gamma_b)^A{}_B$ of
$\gamma_b$. This definition is independent of the induced frames,
i.e.\ independent of the chosen (local) section $E$ in
$S(M,\gb)$. (Once the set of Dirac matrices
$\gamma_0,\ldots,\gamma_3$ is given, $\bgamma$ encodes the
spin-structure at the level of $DM$.)

Moreover, there is an anti-linear isomorphism $DM \to D^*M$ induced by
forming the Dirac adjoint: If $u = u^AE_A$ is a spinor, one can assign
to it a cospinor $u^+ = u^+_BE^B$ with components
\begin{equation}
 u^+_B = \overline{u^A}\gamma_{0AB}
\end{equation}
where $\gamma_{0AB}$ are the matrix elements of $\gamma_0$. This
assignment possesses an inverse (denoted by the same symbol), where a
cospinor $v = v_BE^B$ is mapped to a spinor $v^+$ having components 
$v^{+}{}^A =
\gamma_0{}^{AB}\overline{v_B}$. Again, $\gamma_0{}^{AB}$ are the
matrix elements of $\gamma_0$. The operation of taking the
Dirac adjoint gives rise to anti-linear isomorphisms between
$C^{\infty}(DM)$ and $C^{\infty}(D^*M)$ in the obvious manner.

The metric-induced covariant derivative $\nabla$ on $C^{\infty}(TM)$
induces a covariant derivative, also denoted by $\nabla$, on
$C^{\infty}(DM)$. If $(e_0,\ldots,e_3)$ and $(E_A)_{A = 1}^3$ are
induced by a section $E$ in $S(M,\gb)$ and $f = f^AE_A$ is a local
section in $DM$, then
\begin{equation}
 \nabla f = \nabla_bf^A(e^b \otimes E_A) \in C^{\infty}(T^*M) \otimes
 C^{\infty}(DM) 
\end{equation}
has the frame components
\begin{equation}
 \nabla_bf^A  = \partial_b f^A + \sigma_b{}^A{}_Bf^B\,, \quad {\rm where} \
 \ \ \partial_b f^A = df^A(e_b)\,,\ \ \ \sigma_b{}^A{}_B = -
 \frac{1}{4}\Gamma^a_{bd}\gamma_a{}^A{}_C\gamma^{dC}{}_B\,.
\end{equation}
Here, $df^A$ is the differential of the function $f^A$, and
we read the components of the Dirac matrices on the right hand
side while $\Gamma^a_{bd}$ are Christoffel's connection coefficients,
defined by 
\begin{equation}
 \nabla k = (\partial_b k^a + \Gamma^{a}_{bd}k^d)e^b\otimes e_a
\end{equation}
for $k = k^be_b \in C^{\infty}(TM)$. 

The covariant derivative $\nabla$ can be extended to cospinor fields
and mixed spinor-tensor fields by requiring the Leibniz rule and
commutativity with contractions. Thus, if $h = h_BE^B$ is a cospinor
field, then $\nabla h = \nabla_bh_B e^b \otimes E^B$ has the
components
\begin{equation}
 \nabla_bh_B = \partial_b h_B - h_C \sigma_b{}^C{}_B\,.
\end{equation}
It follows that $\nabla \bgamma = 0$.
\subsection{The Dirac Equation} \label{sect:DirEq}
The Dirac-operator $\dirop$ is a first order differential operator
taking spinor fields to spinor fields, or cospinor fields to cospinor
fields; it is defined as the action of the covariant derivative
followed by contraction with the spinor-tensor $\bgamma$. More
precisely, if $f = f^AE_A \in C^{\infty}(DM)$ and $h = h_BE^B \in
C^{\infty}(D^*M)$, then 
\begin{eqnarray}
\dirop f & = & (\dirop f)^AE_A = \eta^{ab}\gamma_a{}^A{}_B
\nabla_bf^B E_A\,, \\
\dirop h &=& (\dirop h)_BE^B = \eta^{ab}\nabla_bh_C\gamma_a{}^{C}{}_B E^B\,.
\end{eqnarray}
An important property of the Dirac operator is that it commutes with
taking the Dirac adjoint, i.e.\
\begin{equation}
(\dirop f)^+ = \dirop f^+ \quad {\rm and} \quad (\dirop h)^+ = \dirop
h^+
\end{equation}
for $f \in C^{\infty}(DM)$ and $h \in C^{\infty}(D^*M)$\,.

The {\it Dirac equation} is the following first order partial
differential equation for spinor fields $u \in C^{\infty}(DM)$ or for
cospinor fields $v \in C^{\infty}(D^*M)$:
\begin{eqnarray}
 (-i\dirop + m)u & = & 0\,,\label{eq:Direqsp}\\
 (i\dirop + m)v & = & 0\,,
\end{eqnarray}
where $m \ge 0$ is a constant. Then
\begin{equation}
P = (i \dirop + m)(-i \dirop + m)
\end{equation}
is the Lichnerowicz wave operator on spinors or cospinors. It is a second
order wave operator which has metric principal part, and
owing to global hyperbolicity of $(M,\gb)$ this implies
that the Cauchy problem for the corresponding wave equations is
well-posed and that $P$ possesses uniquely determined advanced and
retarded fundamental solutions. As shown in \cite{Dim.D},
this implies that the Dirac operators $-i\dirop + m$ on spinor fields
and $i\dirop + m$ on cospinor fields possess uniquely determined
pairs of advanced$(-)$ and retarded(+) fundamental solutions $\Ssp^{\pm}$ and
$\Sco^{\pm}$, respectively: This means that, for the spinor case,
\begin{equation}
\Ssp^{\pm} : C^{\infty}_0(DM) \to C^{\infty}(DM)
\end{equation}
are continuous linear maps so that
\begin{equation}
(-i\dirop + m)\Ssp^{\pm}u = u = \Ssp^{\pm}(-i\dirop + m)u
\end{equation}
holds for all $u \in C^{\infty}_0(DM)$ and, moreover, ${\rm
  supp}\,\Ssp^{\pm}u$ is contained in the causal future(+)/causal
past$(-)$ of supp\,$u$. (We note that our convention concerning
advanced/re\-tar\-ded fundamental solutions is opposite to that in
\cite{SahlmVer2}.)  The cospinor case is analogous. Then one defines
the retarded-minus-advanced fundamental solutions
\begin{equation}
\Ssp = \Ssp^+ - \Ssp^- \quad {\rm and} \quad \Sco = \Sco^+ -
\Sco^-\,.
\end{equation}

In order to quantize the Dirac field, it is very convenient to
`double' the system by taking pairs of spinor fields and cospinor
fields together, as was done in the references
\cite{Kohler,Kratzert,Hollands}. We give here an equivalent version
which makes contact with the notation used in \cite{SahlmVer2}. To
this end, let us denote $C^{\infty}_0(DM)$ by $\DDsp$ and
$C^{\infty}_0(D^*M)$ by $\DDco$, and define the doubled space $\DDd =
\DDco \oplus \DDsp$. On $\DDd$ we introduce the sesquilinear form   
\begin{equation}
\left( \left[\begin{array}{c} h_1\\ f_1\end{array}\right],
\left[\begin{array}{c} h_2\\ f_2\end{array}\right]\right) =
\langle f_1^+,f_2 \rangle - \langle h_1,h_2^+ \rangle
\end{equation}
for $h_1,h_2 \in \DDco$ and $f_1,f_2 \in \DDsp$, where for $v \in
\DDco$ and $u \in \DDsp$ we employ the dual pairing
\begin{equation}
\langle v,u\rangle = \int_M d\mu_{\gb}(p)\, v_p(u_p)
\end{equation}
with $d\mu_{\gb}$ denoting the canonical 4-volume form induced by the
metric $\gb$ on $M$. This dual pairing also embeds $\DDco$ in the
(topological) dual space $\DDsp'$ of $\DDsp$, and vice versa, embeds
$\DDsp$ in $\DDco'$. The sesquilinear form $(\,.\,,\,.\,)$ is
non-degenerate, but not positive. A useful relation is
\begin{equation}
\label{useful}
 \langle \Sco h,f\rangle = -\langle h,\Ssp f \rangle
\end{equation}
for $h \in \DDco$ and $f \in \DDsp$.

Let us define the conjugate-linear isomorphism $\Gamma: \DDd \to \DDd$,
playing the role of a charge-conjugation, by
\begin{equation}
\Gamma\left( \left[\begin{array}{c} h\\ f\end{array}\right]\right) =
\left[\begin{array}{c} f^+\\ h^+\end{array}\right]  \,.
\end{equation}
Then one finds that $\Gamma$ is a skew-conjugation with respect to
$(\,.\,,\,.\,)$, namely it holds that
\begin{equation}
(\Gamma F_1,\Gamma F_2) = - (F_2,F_1) \quad \forall\ F_1,F_2 \in
\DDd\,.
\end{equation}
Now we introduce the following `doubled' operators on $\DDd$:
\begin{eqnarray}
D_\rhd &=& \left(\begin{array}{cc}
          - \dirop + im & 0 \\
            0 & \dirop + im
          \end{array}\right)\,,\ \ \ \
D_\lhd = \left(\begin{array}{cc}
          - \dirop - im & 0 \\
            0 & \dirop - im
          \end{array}\right)\,,\\
& & \quad \quad \quad S_\lhd = \left(\begin{array}{cc}
           i\Sco & 0 \\
            0 & i\Ssp
          \end{array}\right)\,.
\end{eqnarray}
Then it holds that
\begin{equation}
D_\lhd D_\rhd = D_\rhd D_\lhd = P_{\rm double} =   \left(\begin{array}{cc}
          P_{\rm cosp}& 0 \\
            0 & P_{\rm sp}
          \end{array}\right)
\end{equation}
where $P_{...}$ denotes the Lichnerowicz wave operator on spinors and
cospinors, respectively; moreover, one finds that $\Gamma$ commutes
with $P_{\rm double}$ and
\begin{eqnarray}
\label{GammaRel}
\Gamma D_\lhd &=& - D_\lhd \Gamma\,, \quad \Gamma D_\rhd = - D_\rhd
\Gamma, \\
\label{DRel}
{}\hspace*{-0.9cm}(D_\lhd F_1,F_2) &=& -(F_1,D_\lhd F_2)\,, \quad
(D_\rhd,F_1,F_2) = - 
(F_1,D_\rhd F_2)\ \ \forall\ F_1,F_2 \in \DDd\,.
\end{eqnarray}
One may also check that $S_\lhd^{\pm}$ (defined in obvious analogy to
$S_\lhd$) are the retarded(+)/ad\-van\-ced$(-)$ fundamental solutions for
the operator $D_\rhd$; consequently 
\begin{equation}
\label{DirSol}
 D_\rhd S_\lhd = S_\lhd D_\rhd = 0\,.
\end{equation}
Furthermore, from \eqref{GammaRel} one can see that
\begin{equation} \label{GS}
 \Gamma S_\lhd = -
S_\lhd \Gamma\,.
\end{equation}
 This entails that $\Gamma$ is a complex conjugation
for the sesquilinear form 
\begin{equation}
(F_1,F_2)_S = (S_\lhd F_1,F_2)\,, \quad F_1,F_2 \in \DDd\,,
\end{equation}
so that
\begin{equation}
 (\Gamma F_1,\Gamma F_2)_S = (F_2,F_1)_S\ \ \ \forall\ F_1,F_2 \in
 \DDd\,.
\end{equation}
On the other hand one can see that (cf.\ \eqref{useful})
\begin{equation}
 \left( \left[\begin{array}{c} h_1\\ f_1\end{array}\right],
\left[\begin{array}{c} h_2\\ f_2\end{array}\right]\right)_S = 
-i\langle  f_1^+,\Ssp f_2\rangle + i \langle\Sco h_2, h_1^+\rangle
\end{equation} 
and this implies by Prop.\ 2.2 in \cite{Dim.D} that
$(\,.\,,\,.\,)_S$ is positive-semidefinite, $(F,F)_S \ge 0$.

Now we introduce the quotient space $\DDd/{\rm ker}\,S_\lhd$ and
denote by $\HH$ its completion with respect to $(\,.\,,\,.\,)_S$. The
conjugation $\Gamma$ induces by \eqref{GS} a conjugation on $\HH$
which will again be denoted by $\Gamma$. Hence, we have derived from
the doubled Dirac equation a complex Hilbert-space $\HH$ (with scalar
product $(\,.\,,\,.\,)_S$) together with a complex conjugation
$\Gamma$. The system can be quantized, following Araki \cite{Araki},
by assigning to these data the 
algebra of canonical anti-commutation relations
CAR$(\HH,\Gamma)$. This is the unique $C^*$-algebra with unit $\II$
which is generated by a family $\{B({\bf v}): {\bf v} \in \HH\}$
subject to the relations:
\begin{itemize}
\item[(i)] ${\bf v} \mapsto B({\bf v})$ is $\CC$-linear,
\item[(ii)] $B(\Gamma{\bf v}) = B({\bf v})^*$\,,
\item[(iii)] $B({\bf v})^*B({\bf w}) + B({\bf w})B({\bf v})^* = ({\bf
    v},{\bf w})_S \cdot \II$\,.
\end{itemize}
Now let $q: \DDd \to \DDd/{\rm ker}\,S_\lhd$ denote the quotient map,
then we define the quantized Dirac field as the linear map which
assigns to each $h \in \DDco$ the element
\begin{equation}
 \Psi(h) = B\left(q\left[\begin{array}{c} h\\
       0\end{array}\right]\right)
\end{equation}
in CAR$(\HH,\Gamma)$. The adjoint spinor field will be defined by
\begin{equation}
 \Psi^+(f) = B\left(q\left[\begin{array}{c} 0\\
       f\end{array}\right]\right)\,, \quad f \in \DDsp\,.
\end{equation}
As a consequence of (iii), the field and its adjoint satisfy the
anti-commutation relations
\begin{equation}
 \Psi(h)\Psi^+(f) + \Psi^+(f)\Psi(h) = -i\langle h,\Ssp f \rangle\,.
\end{equation}
We also note that, owing to \eqref{DirSol}, $B(q(D_\rhd F)) = 0$, and
this entails
\begin{equation}
\Psi((i \dirop + m)h) = 0 \quad {\rm and} \quad \Psi^+((-i\dirop +
m)f) = 0
\end{equation}
for all $h \in \DDco$ and $f \in \DDsp$.
\\[10pt]
{\bf Remarks. } (i) $\Psi$ acts linearly on cospinors and fulfills, in
the sense of distributions, the equation $(-i\dirop + m)\Psi = 0$,
therefore the map $h \mapsto \Psi(h)$ is regarded as a spinor
field. Similarly, $\Psi^+$ acts linearly on spinor fields and fulfills
in distributional sense $(i\dirop + m)\Psi^+ = 0$; hence it is viewed as
a cospinor field.
\\[6pt]
(ii) $\Psi$ and $\Psi^+$ are $C^*$-valued distributions since e.g.\
$2||\Psi^+(f)||^2 =- i\langle f^+,\Ssp f\rangle$ and $\DDsp \otimes \DDsp
\owns f_1 \otimes f_2 \mapsto -i\langle f^+_1,\Ssp f_2 \rangle$
is continuous (with respect to the usual test-function topology).
\\[6pt]
(iii) We briefly comment on the case where $\gamma_0,\ldots,\gamma_3$
belong to a Majorana representation (as considered in
\cite{SahlmVer2}) and one wishes to quantize the Majorana field. 
In that situation, $\DDco$ (and similarly, $\DDsp$) carries an
`intrinsic' charge conjugation $\Gamma$ given by $v= v_BE^B \mapsto
\overline{v_B}E^B$ in any frame. Then $\Gamma$ is a skew-conjugation
for the sesquilinear form
\begin{equation}
 (h,h') = \int_M d\mu_{\gb}(p)\, h'_p(h_p^+)
\end{equation}
on $\DDco$. Upon defining 
 $D_\rhd = \dirop - im$, $D_\lhd = \dirop + im$ and $S_\lhd =
i\Sco$, one obtains similar relations as before. Then
$\HH$ arises as completion of $\DDco/{\rm ker}\,S_\lhd$ and $(h,h')_S
= (S_\lhd h,h')$. 
The field operators then simplify to
\begin{equation}
 \Psi(h) = B(q(h));\qquad \Psi^+(f)=\Psi(f^+)^*=\Psi(\Gamma f^+)
\end{equation}
for $h \in \DDco$, $f\in\DDsp$. In other words, the Majorana field may be quantized
without doubling the classical system.
Instead of starting with $\DDco$ one can likewise
consider $\DDsp$; this has been done in \cite{SahlmVer2}. 
\subsection{Hadamard states} \label{sect:Hadstate}
We recall that a state on a $C^*$-algebra ${\cal C}$ is a linear
functional $\omega: {\cal C} \to \CC$ fulfilling $\omega(\II) = 1$ and
$\omega(A^*A) \ge 0$ for all $A \in {\cal C}$. For the purpose of the
present work, it is sufficient to focus on the two-point functions
$\omega_2$ of states $\omega$ on CAR$(\HH,\Gamma)$. The two-point
function $\omega_2$ of $\omega$ is an element in $(\DDd \otimes
\DDd)'$ given by
\begin{equation}
\omega_2(F_1 \otimes F_2) = \omega(B(q(F_1))B(q(F_2)))\,, \quad
F_1,F_2 \in \DDd\,.
\end{equation}
It was shown in \cite{Araki} that there is for each state $\omega$ on
CAR$(\HH,\Gamma)$ a linear operator $Q$ on $\HH$ with the properties:
\begin{itemize}
\item[(I)] $0 \le Q^* = Q \le 1$\,,
\item[(II)] $Q + \Gamma Q \Gamma = \II$\,,
\item[(III)] $\omega_2(F_1 \otimes F_2) = (\Gamma q(F_1),Q q(F_2))_S$
  \quad for all $F_1,F_2 \in  \DDd$.
\end{itemize}
Conversely, each linear operator $Q$ with the properties (I) and (II)
determines by (III) the two-point function $\omega_2$ of some state
$\omega$ on CAR$(\HH,\Gamma)$, a so-called {\it quasifree state},
determined by the two-point function $\omega_2$ (see \cite{Araki} for
discussion). Such a quasifree state $\omega$ is pure (and then often
called a Fock-state) if and only if $Q$ is a projection, i.e.\ $Q^2 =
Q$. The operator $Q$ with the properties (I), (II) and (III) above
will be called the {\it operator labelling the quasifree state $\omega$}.

The stress-energy tensor is defined using the two-point functions of a
particular class of states, the {\it Hadamard states}. One says that a
state $\omega$ on CAR$(\HH,\Gamma)$ is a Hadamard state if we may write
\begin{equation}
\omega_2(F_1 \otimes F_2) = w(D_\lhd F_1 \otimes F_2)
\end{equation}
for some distribution $w \in (\DDd \otimes \DDd)'$ of
Hadamard form for the doubled wave-operator $P_{\rm double}$ on
$\DDd$; the definition of a Hadamard form for such a wave-operator has
been given in \cite{SahlmVer2} (cf.\ also
\cite{Kohler,Kratzert,Hollands,NO}). This definition entails that the
difference between the two-point functions of two Hadamard states is smooth. 
For the purposes of this
work, we will also need the characterization of Hadamard states in terms of
properties of the wave-front set WF$(\omega_2)$ that appears in
\cite{Kratzert,Hollands,SahlmVer2}, following a line of argument given
in \cite{Rad1} for the scalar case. The relevant statement, proven in
the references just stated, is: A state
$\omega$ on CAR$(\HH,\Gamma)$ is a Hadamard state if and only if the
wave-front set of its two-point function $\omega_2$ satisfies the
relation
\begin{equation}
\label{muSC1}
\WF(\omega_2) = 
\left\{(p,\xi;p',-\xi')\in\dot{T}^*(M\times M) \mid (p,\xi) \sim
(p',\xi');\quad \xi\in \Nn_{p}^+ \right\}\,,
\end{equation}
where $\dot{T}^*(M \times M)$ is the cotangent bundle over $M \times
M$ without the zero-section, $(p,\xi) \sim (p',\xi')$ means that there
is a lightlike geodesic connecting the points $p$ and $p'$ in $M$ and
to which $\xi$ and $\xi'$ are co-tangent, and $\Nn_{p}^+$ is the set
of all future-directed null covectors at $p$. (We remark that in
\cite{SahlmVer2} the opposite sign convention for the
Fourier transform was chosen, leading to the opposite constraint $\xi
\in \Nn_p^-$ (i.e., past-directed null covectors) in that reference 
compared to \eqref{muSC1}.)

At this point we very briefly recall the definition of the wave-front
set of a distribution~\cite{Hormander1}. For a distribution $t \in 
  \DD'(\RR^n)$, a point $(x,k) \in \RR^n \times (\RR^n \backslash
\{0\})$ is called a regular directed point of $t$ if there exists
$\chi \in \DD(\RR^n)$ with $\chi(x) \ne 0$ and a conic open
neighbourhood $C$ of $k$ in $\RR^n \backslash \{0\}$ such that 
\begin{equation}
 \sup_{k' \in C}\,(1 + |k'|)^N |\widehat{\chi t}(k')| < \infty \quad
 \forall \ N \in \NN\,.
\end{equation}
(If this holds we will say that $\widehat{\chi t}$ is of {\rm rapid
decay} in $C$.)
The complement in $\RR^n \times(\RR^n \backslash \{0\})$ of the set of
all regular directed points of $t$ is called the wave-front set $\WF
(t)$ of $t$. Given a scalar distribution $\tau$ on a manifold $M$, one
says that a non-zero covector $(p,\xi) \in \dot{T}^*(M)$ is in $\WF
(\tau)$ if there is a coordinate chart $(U,\kappa)$ around $p \in M$ so
that $(\kappa(p),{}^t\!(\kappa^{-1})'\xi) \in \WF (\tau\circ\kappa^{-1})$
where (as discussed at the end of Sec.~\ref{sect:intro})
$\tau\circ\kappa^{-1}\in\DD'(\kappa(U))$ is a distribution on the chart range of
$\kappa$. This definition of $\WF(\tau)$ is independent of the choice
of the chart $\kappa$. We refer the reader to \cite{Hormander1} for
further discussion of the properties of wave-front set of distributions
on manifolds. For the case that $\tau$ is a distribution on test-sections
of a vector bundle, e.g.\ defined on $\DDd$, $\tau$ can be viewed, via
(local) trivializations of the bundle, as collection $(\tau^A{}_B)_{A,B}$
of scalar distributions, and then $\WF(\tau)$ is defined as the union
of $\WF(\tau^A{}_B)$ over all components $A,B$. It is not difficult to
see that this definition is independent of the chosen (local) trivialization.
 
Now let $\omega_2$ be the two-point function of a Hadamard state
$\omega$ on
CAR$(\HH,\Gamma)$, and let $Q$ be the corresponding operator on $\HH$
with the properties (I),(II) and (III)
above. We will use the notation 
\begin{equation}
 Q^{\Gamma} = \Gamma Q \Gamma
\end{equation}
 for the
`charge conjugate' of $Q$, and we
will adopt this notation also for other operators on $\HH$. In
studying the stress-energy tensor, we will be particularly interested
in the following distributions on $\DDsp \otimes \DDco$ associated with
$\omega$ (respectively, with $Q$), which we will also refer to as
two-point functions:
\begin{equation}
 \omega_Q(f\otimes h) = \omega(\Psi^+(f)\Psi(h))\,, \quad
 \omega^{\Gamma}_Q(f\otimes h) = \omega(\Psi(h)\Psi^+(f))
\label{eq:omegaQdef}
\end{equation}
where $f \in \DDsp$ and $h \in \DDco$. Note that $\omega^{\Gamma}_Q =
\omega_{Q^{\Gamma}}$. As a consequence of the constraint on the
wave-front set~\eqref{muSC1} for Hadamard states, which is also called
{\it microlocal spectrum condition}, one finds that the following
microlocal spectrum condition holds for $\omega_Q$ and
$\omega_Q^\Gamma$,
\begin{equation}
\WF(\omega_Q^\sharp) = 
\left\{(p,\xi;p',-\xi')\in\dot{T}^*(M\times M) \mid (p,\xi)\sim
(p',\xi');\quad \xi\in \Nn_{p}^\sharp\right\}.
\label{eq:uSC2}
\end{equation}
Here, and below, we use $\sharp$ and $\flat$ to denote
either the presence or absence of a $\Gamma$ in the following
context-dependent way:
\begin{center}
\begin{tabular}{cccc} 
$Q^\sharp$ & $\omega_{Q}^\sharp$ & $\Nn_p^\sharp$ & $\RR^\sharp$
\\[2pt] \hline\hline
$Q$ & $\omega_{Q}$ & $\Nn_p^+$ & $\RR^+$ \\[2pt]
$Q^\Gamma$ & $\omega_{Q}^\Gamma$ & $\Nn_p^-$ & $\RR^-$ 
\end{tabular}\,\,.
\end{center}
To avoid confusion we will sometimes use $\cdot$ as a placeholder to
indicate the absence of a $\Gamma$. Thus $Q^\cdot=Q$, for example.
We note that the microlocal spectrum condition in the
form~(\ref{eq:uSC2}) has been
proved directly for quasifree Hadamard states in \cite{Kratzert,Hollands}.

Let us note that~(\ref{eq:omegaQdef}) may be written 
\begin{equation}
 \omega_Q(f \otimes h) = 
\left(\Gamma q \left[\begin{array}{c} 0\\ f\end{array}\right],
Q q \left[\begin{array}{c} h\\ 0\end{array}\right]\right)_S \,;
\label{eq:omegaQgen}
\end{equation}
as a slight abuse of notation we will use this relation to {\em define}
$\omega_Q$ for general bounded operators $Q$ on $\HH$ and refer to
$\omega_Q$ as the {\em two-point function labelled by $Q$}. 
(Of course $\omega_Q$ is not in general the two-point function of a state.)
We will also denote by $\Had(\HH,\Gamma)$ the class of operators which
obey properties (I) and (II) and such that $\omega_Q^\sharp$
obeys~(\ref{eq:uSC2}). Thus $\Had(\HH,\Gamma)$ parametrizes the
quasifree Hadamard states on ${\rm CAR}(\HH,\Gamma)$.

Our proof of Thm.\ 4.1 below relies on the existence of pure,
quasifree Hadamard 
states for the Dirac field. It seems that this has never been
established in the literature in full detail, therefore we sketch here
how such states may be constructed by adapting an argument employed by
Fulling, Narcowich and Wald \cite{FNW} for the case of the free scalar
field to the Dirac field. The first step is to show that there exists
a pure, quasifree Hadamard state for the Dirac field for ultrastatic
$(M,\gb)$. In fact, if $(M,\gb)$ is ultrastatic, then
it may be endowed with a suitable spin structure so 
that the ultrastatic time shifts give
rise to a continuous unitary group $U_t$ $(t \in \RR)$ on $\HH$
leaving the scalar product $(\,.\,,\,.\,)_S$ invariant and fulfilling
$\Gamma U_t = U_t \Gamma$. Moreover, if the mass parameter $m$
appearing in the Dirac equation is strictly positive, then the
spectrum of the self-adjoint generator of $U_t$ is bounded away from
zero. Theorem 2 in \cite{Araki}, or the results in \cite{Weinless},
thus show that there is a pure quasifree state $\omega_0$ on the CAR
algebra of the Dirac field on ultrastatic $(M,\gb)$ which is a ground
state for the $C^*$-dynamics induced by $U_t$; the projection $P_0$
labelling $\omega_0$ is
the projection onto the positive spectral subspace of the unitary
group $U_t$. Since $\omega_0$ is a ground state, it fulfills the
microlocal spectrum condition \cite{SahlmVer1} and hence is a Hadamard
state \cite{SahlmVer2}. In a second step, one uses a technique
developed in \cite{FNW} which allows one to view a neighbourhood of a
Cauchy surface of any given globally hyperbolic spacetime as being
isometrically embedded in a globally hyperbolic spacetime that has an
ultrastatic part (with suitable spin structure as above) in its past.
By the uniqueness of the Cauchy problem
and the ``propagation of Hadamard form'' under the dynamics of the
Dirac equation \cite{SahlmVer2}, any pure, quasifree Hadamard state
prescribed on the ultrastatic part of the spacetime (e.g.\ $\omega_0$)
induces a pure, quasifree Hadamard state everywhere on the spacetime,
in particular on the embedded neighbourhood of the Cauchy surface of
the initially given globally hyperbolic spacetime. Using the same
argument once more, a pure, quasifree Hadamard state for the Dirac
field is thereby induced on any given globally hyperbolic
spacetime. The mass parameter $m$ may be allowed to be variable over
spacetime in this process without affecting the Hadamard form, so that
one obtains a pure, quasifree Hadamard state of the Dirac field on any
globally hyperbolic spacetime for any $m \ge 0$. The argument just sketched
implicitly also shows that there exists an abundance of
quasifree Hadamard states. 

\section{A point-split energy density}
\label{sect:psed}

For the remainder of this paper, we will assume that $(M,\gb)$ is
globally hyperbolic, orientable and time orientable, with spin structure
$(S(M,\gb),\psi)$ and that the Dirac matrices $\gamma_a$ belong to a 
standard representation. 

Let $\gamma:\RR\to M$ be a smooth timelike curve in $(M,\gb)$, parametrized by
its proper time, along which we wish to establish a QWEI. The starting
point is the construction of a normal ordered energy density on
$\gamma$, which is accomplished as follows. We first claim that there
exists a tubular neighbourhood $C_\gamma$ of $\gamma$ and a local
section $E$ of $S(M,\gb)$ over $C_\gamma$ such that the induced tetrad
field $(e_0,\ldots,e_3)=\psi\circ E$ satisfies $e_0|_\gamma=u$, where $u= 
\dot{\gamma}$ is the velocity of $\gamma$. To see this, choose any
locally finite open cover $\{U_j\mid j\in\ZZ\}$ of $\gamma$ by charts
$U_j$ such that
\renewcommand{\theenumi}{(\roman{enumi})}
\begin{enumerate}
\item $U_j\cap U_k=\emptyset$ unless $|j-k|\le 1$;
\item $U_j\cap U_{j+1}$ is contractable for each $j\in\ZZ$;
\item $U_j\cap U_k \cap U_l=\emptyset$ if $j,k,l$ are distinct.
\end{enumerate}
The existence of such a cover follows from global hyperbolicity of
$(M,\gb)$ since $\gamma$ is timelike. Now extend $u$ to a smooth
timelike unit vector field $\widetilde{u}$ on some tubular neighbourhood
$C_\gamma$ of $\gamma$, so that $C_\gamma\subset \bigcup_j U_j$. Choose any
tetrad $(e_0',\ldots,e_3')$ on $C_\gamma$. Then we may obtain a tetrad 
$(e_0,\ldots,e_3)$ with $e_0=\widetilde{u}$ by applying a unique boost in
$\Lorpror$ at each point (whose parameters are given by the components
of $\widetilde{u}$ with respect to $e_a'$, and therefore vary smoothly).
This tetrad lifts smoothly to $S(M,\gb)$ in each $U_j\cap C_\gamma$ and may be
patched together along $C_\gamma$ to obtain the required section $E$ 
by virtue of properties (i), (ii) and~(iii).\footnote{Of course, there
are exactly two such sections.} 

Next, we choose smooth spinor fields $v_A$ ($A=1,\ldots,4$) in
$C_\gamma$, such that
\begin{equation}
\delta^{AB} v_A \otimes v_B^+ = \gamma_0\,. \label{eq:g0dec}
\end{equation}
This is easily satisfied by taking $v_A=E_A$, where $E_A$ is the spin
frame induced by $E$; however, it will be
convenient to make a slightly different choice when considering the
Majorana field. The changes relevant for Majorana fields 
will be described in Sec.~\ref{sect:Maj}. 

With respect to the frame $e_a$ the Dirac stress-energy tensor is
\begin{equation}
T_{ab} = \frac{i}{2} \left(\psi^+ \gamma_{(a}\nabla_{b)} \psi -
(\nabla_{(a}\psi^+)\gamma_{b)} \psi\right)\,,
\end{equation}
which is manifestly symmetric, and is conserved provided $\psi$ obeys the
Dirac equation~(\ref{eq:Direqsp}). In particular, $T_{00}(\gamma(\tau))$
is the energy density measured by an observer with worldline $\gamma$ at
proper time $\tau$. 

We may use~(\ref{eq:g0dec}) to define a bi-scalar point-split energy density
\begin{equation}
\TT(x,y) = 
\delta^{AB}\frac{i}{2}\left((\psi^+ v_A)(x)  (v_B^+ e_0\cdot\nabla\psi)(y) 
-  ([e_0\cdot\nabla\psi^+]v_A)(x) (v_B^+ \psi)(y)
\right)
\end{equation}
with the property that $\TT(x,x)=T_{00}(x)$. Integrating by parts, $\TT$
becomes a scalar bi-distribution $\TT\in(\DD(M)\otimes\DD(M))'$
\begin{equation}
\label{iii4}
\TT(f\otimes g) = \delta^{AB}\frac{i}{2}\left(
 \psi^+(\nabla\cdot [e_0 v_A f])\psi(v_B^+ g) 
-\psi^+( v_A f) \psi(\nabla\cdot[e_0 v_B^+ g]) 
\right)\,.
\end{equation}
Here (and below) the notation $\nabla\cdot [e_0 v]$ denotes minus the
distributional dual of $e_0\cdot\nabla$, applied to the test function
or (co)spinor $v$. Thus,
\begin{equation}
\nabla\cdot [e_0 v] = v\nabla\cdot e_0 + e_0\cdot\nabla v
\end{equation}
where, with respect to local coordinates $(x^{\mu})$, $\nabla \cdot
e_0 = \nabla_{\mu}e_0^{\mu}$ and $e_0 \cdot \nabla v = e_0^{\mu}\nabla_{\mu}v$.
Upon quantization, we obtain the algebra-valued bi-distribution $T$
given by
\begin{equation}
\label{iii6}
T(f\otimes g) = \delta^{AB}\frac{i}{2}\left(
\Psi^+(\nabla\cdot [e_0 v_A f]) \Psi(v_B^+ g) 
-\Psi^+(v_A f) \Psi(\nabla\cdot[e_0 v_B^+ g])
\right)\,.
\end{equation}

Given a state $\omega$ we will now use the same symbol to denote its
two-point function $\omega(f\otimes g)=\omega(\Psi^+(f)\Psi(g))$ and also set
$v_{AB}=v_A\otimes v_B^+$. Thus $v_{AB}\omega$ will denote the matrix of
scalar bi-distributions
\begin{equation}
v_{AB}\omega(f\otimes g) = \omega(\Psi^+(v_A f)\Psi(v_B^+g))\,.
\end{equation}
The formulae $\nabla\cdot(e_0 v_A f) = v_A \nabla\cdot(e_0  f) + 
\sigma_{0}{}^C{}_A f v_C$ and $\nabla\cdot(e_0 v_B^+ g) = v_B^+ \nabla\cdot(e_0 
g)+ \overline{\sigma_{0}{}^C{}_B} g v_C^+$ now allow us to write the
expectation value 
$\langle T\rangle_\omega$ of $T$ in state $\omega$ as 
\begin{equation}
\langle T\rangle_\omega =\LL^{AB} v_{AB}\omega\,,
\end{equation}
where
\begin{equation}
\LL^{AB}= \frac{1}{2}
\left(\II\otimes ie_0\cdot\nabla-ie_0\cdot\nabla\otimes\II \right) \delta^{AB}
+ \frac{1}{2}\Theta^{AB} \,,
\end{equation}
and
\begin{equation}
\Theta^{AB} =
i [\delta^{CB}\sigma_{0}{}^{A}{}_C\otimes\II -\II\otimes
\delta^{AC}\overline{\sigma_{0}{}^B{}_C}]\,.
\end{equation}

If a reference Hadamard state $\omega_0$ is now specified, we may define the
normal ordered point-split energy density (with respect to $\omega_0$)
by
\begin{equation}
\langle :T:\rangle_\omega = \langle T\rangle_\omega - \langle T\rangle_{\omega_0}\,.
\end{equation}
This may also be written 
\begin{equation}
\langle
:T:\rangle_\omega=\LL^{AB}v_{AB}:\omega:\,,
\end{equation}
where $:\omega:=\omega-\omega_0$ is the normal ordered two-point function.
Since $:\omega:$ is smooth for Hadamard $\omega$, $\langle
:T:\rangle_\omega$ is also smooth.  Accordingly, the `coincidence
limit' (i.e., the restriction of $\langle :T:\rangle_\omega$ to the
diagonal) is well defined and yields the normal ordered energy density
near $\gamma$. We denote the energy density along $\gamma$ by
\begin{equation}
\langle :\rho:\rangle_\omega (\tau) = \langle :T:\rangle_\omega(\gamma(\tau),\gamma(\tau))\,.
\label{eq:rho}
\end{equation}
Note also that $\langle:T:\rangle_\omega(f\otimes g)$ is symmetric in
$f$ and $g$ by virtue of the CAR's. 

It will be convenient to regard $\langle :\rho:\rangle_\omega$ as the
diagonal of the pull-back $\gamma_2^* \langle :T:\rangle_\omega$ where
$\gamma_2(\tau,\tau')=(\gamma(\tau),\gamma(\tau'))$. In turn, $\gamma_2^* \langle
:T:\rangle_\omega$ may be written as the action of a differential operator on the
pulled-back normal ordered two-point function:
\begin{equation}
\gamma_2^* \langle :T:\rangle_\omega = \frac{1}{2}\left[(\II\otimes
D-D\otimes\II)\delta^{AB} +\gamma_2^* \Theta^{AB} \right]\gamma_2^* v_{AB}:\omega: \,,
\label{eq:g2T}
\end{equation}
where $D=id/d\tau$ (strictly speaking, $D$ should be regarded as the 
distributional dual of $-id/d\tau$).

\section{Main argument}\label{sect:main}

We now come to the proof of the QWEI for Dirac fields. In the following,
$(M,\gb)$ is assumed to satisfy the hypotheses stated at the beginning of the
previous section.

\begin{Thm} \label{thm:main} 
Let $\gamma:\RR\to M$ be a smooth timelike curve in $(M,\gb)$
parametrized by its proper time. Let
$\omega_0$ be a Hadamard state of the Dirac field on $(M,\gb)$. Define
the normal ordered energy density $\langle:\rho:\rangle_\omega$ by~(\ref{eq:rho})
with respect to the reference state $\omega_0$. 
Then for any weight $f$ belonging to $\WW$ (defined in Eq.~(\ref{eq:weights}))
\begin{equation}
\inf_{\omega} \int d\tau\,\langle:\rho:\rangle_\omega (\tau) f(\tau)
>-\infty,
\label{eq:QI}
\end{equation}
where the infimum is taken over all Hadamard states $\omega$. 
That is, there exists a quantum weak energy inequality for the Dirac
field. 
\end{Thm}
{\noindent\bf Remarks.} (i) If the reference state $\omega_0$ is changed,
$\langle:\rho:\rangle_\omega$ is modified by a smooth function which is
independent of $\omega$. Thus we may assume without loss of generality
that $\omega_0$ is pure and quasifree. Exactly the same argument entails
that~(\ref{eq:QI}) holds if we replace the normal ordered energy density
by the renormalised energy density.  \\[6pt]
{\noindent (ii)} Perhaps surprisingly, the class $\WW$ (of squares of
real-valued 
$\CoinX{\RR}$ functions) does not coincide with the class of nonnegative
smooth compactly supported functions. In fact,
Glaeser~\cite{Glaeser}\footnote{We 
are grateful to S.P.\ Eveson and P.J.\ Bushell for help in locating
this reference.} has 
constructed an example of a $C^\infty$ nonnegative function $f$, vanishing only
at the origin, so that $\frac{d^2}{dx^2}\sqrt{f(x)}$ diverges as $x
\to 0$. The delicacy of this
point resides in the behaviour of $f$ at
zeros of infinite order. It is not clear whether the restriction to
weights in $\WW$ is purely a technical limitation 
of our proof, or whether QWEIs should be understood as quadratic form
results (cf.~\cite{FewsterTeo2}).

{\noindent\em Proof:} It is sufficient to prove~(\ref{eq:QI}) for
arbitrary $f\in\CoinX{I} \cap \WW$ where $I\subset\RR$ is an arbitrary 
open interval with compact closure. 
To this end, choose $\eta\in\CoinX{M}$ such that $\eta$ equals
unity on a neighbourhood of $\gamma(I)$. It is easy to see that
$\langle:\rho:\rangle_\omega$ is unaltered on $I$ if we replace
$v_{AB}:\omega:$ in~(\ref{eq:g2T}) by the compactly supported distribution
$u_{AB}:\omega:$, where 
\begin{equation}
u_{AB} = \eta v_A\otimes \eta v_B^+\,.
\label{eq:uAB}
\end{equation}

Applying the formula
\begin{equation}
\int d\tau\, F(\tau,\tau)\varphi(\tau) = \int \frac{d\lambda\,d\lambda'}{(2\pi)^2}
\widehat{F}(-\lambda,\lambda') \widehat{\varphi}(\lambda-\lambda')\,,
\end{equation}
which is valid for $F\in\CoinX{\RR^2}$ and $\varphi\in\CoinX{\RR}$, one
may show that
\begin{equation}
\cI= \int d\tau\,\langle:\rho:\rangle_\omega(\tau) f(\tau)
= \int d\lambda\,d\lambda' J^{AB}(\lambda,\lambda')
 W_{AB}^{(\omega)}(\lambda,\lambda') 
\label{eq:Idef}
\end{equation}
where
\begin{equation}
W_{AB}^{(\omega)}(\lambda,\lambda') = 
\left[\gamma_2^* u_{AB}:\omega:
\right]^\wedge(-\lambda,\lambda')\,,
\end{equation}
and we have also written
\begin{equation}
J^{AB}(\lambda,\lambda')=\frac{1}{8\pi^2}\left\{
(\lambda+\lambda')\widehat{f}(\lambda-\lambda')\delta^{AB}
+[\theta^{AB}f]^\wedge(\lambda-\lambda') \right\}\,,
\end{equation}
where
\begin{equation}
\theta^{AB}(\tau)=\gamma_2^*\Theta^{AB}(\tau,\tau) =
i\left(\delta^{CB}\sigma_0{}^A{}_C|_{\gamma(\tau)} -
\delta^{AC}\overline{\sigma_0{}^B{}_C|_{\gamma(\tau)}}\right)
\end{equation}
is clearly hermitian ($\overline{\theta^{BA}(\tau)}=\theta^{AB}(\tau)$).
It follows that $J^{AB}$ is a hermitian matrix kernel, i.e., 
$J^{AB}(\lambda,\lambda')=\overline{J^{BA}(\lambda',\lambda)}$.

Note that $J^{AB}$ is state-independent, while $W_{AB}^{(\omega)}$ 
contains all the dependence on the state of interest $\omega$ and the
reference state $\omega_0$. We also note that 
$J^{AB}(\lambda,\lambda')$ decays rapidly away
from the diagonal in $\RR^2$.

Assuming without loss that $\omega_0$ is pure and quasifree, it must be
labelled by some projection $P$ on
$\HH$. Since the stress-energy tensor is defined in terms of the two-point
function, it is enough to establish~(\ref{eq:QI}) when the infimum
is taken over quasifree Hadamard states, whose two-point functions are of
the form $\omega_Q$ for $Q\in\Had(\HH,\Gamma)$ as discussed in
Sec.~\ref{sect:Hadstate}. The normal ordered two-point function
$:\omega_Q:=\omega_Q-\omega_P$ is labelled by $Q-P$, i.e., 
$:\omega_Q:=\omega_{Q-P}$ in the spirit of the remarks
following~(\ref{eq:omegaQgen}). Now 
\begin{equation}
Q-P = -PQ^\Gamma P + P^\Gamma Q P + P Q P^\Gamma + P^\Gamma Q P^\Gamma\,,
\label{eq:decomp}
\end{equation}
and this induces a decomposition of $:\omega_Q:$ into four pieces
\begin{equation}
:\omega_Q: = - \omega^{\cdot\,\Gamma\,\cdot} + 
\omega^{\Gamma\,\cdot\,\cdot} + \omega^{\cdot\,\cdot\,\Gamma} + 
\omega^{\Gamma\,\cdot\,\Gamma}\,,
\end{equation}
where $\omega^{\sharp\,\natural\,\flat}$ is the two-point function
labelled by $P^\sharp Q^\natural P^\flat$; that is,
\begin{equation}
 \omega^{\sharp\,\natural\,\flat}(f \otimes h) = 
\left(\Gamma q \left[\begin{array}{c} 0\\ f\end{array}\right],P^\sharp
  Q^\natural P^\flat q
\left[\begin{array}{c} h\\ 0\end{array}\right]\right)_S 
\label{eq:omsnf}
\end{equation}
for $f \in \DDsp$ and $h \in \DDco$. In Sec.~\ref{sect:pfdineq} we
will show that each $\omega^{\sharp\,\natural\,\flat}$ may be pulled
back to $\RR^2$ by $\gamma_2$, allowing us to write
\begin{equation}
W_{AB}^{(\omega_Q)} = - W_{AB}^{\cdot\,\Gamma\,\cdot} + 
W_{AB}^{\Gamma\,\cdot\,\cdot} + W_{AB}^{\cdot\,\cdot\,\Gamma} + 
W_{AB}^{\Gamma\,\cdot\,\Gamma}\,,
\label{eq:dec}
\end{equation}
where
\begin{equation}
W_{AB}^{\sharp\,\natural\,\flat}(\lambda,\lambda') = 
\left[\gamma_2^* u_{AB}\omega^{\sharp\,\natural\,\flat}
\right]^\wedge(-\lambda,\lambda')\,.
\label{eq:WABsnf}
\end{equation}
Furthermore, the following $Q$-independent bounds will be established in 
Sec.~\ref{sect:pfdineq}.
\begin{Lem} \label{lem:dineq}
For any $Q\in\Had(\HH,\Gamma)$, 
\begin{equation}
\left| W_{AB}^{\sharp\,\natural\,\flat}(\lambda,\lambda')\right| \le 
X_{AB}^{\sharp\,\flat}(\lambda,\lambda') \qquad\forall (\lambda,\lambda')\in\RR^2\,,
\label{eq:dineq}
\end{equation}
where $X_{AB}$ is independent of $Q$ and is defined in terms of the
reference two-point function $\omega_P$ by
\begin{equation}
X_{AB}^{\sharp\,\flat}(\lambda,\lambda') = Y_{A}^\sharp(\lambda)
Y_{B}^\flat(\lambda')\,,
\end{equation}
where $Y_A^\sharp(\lambda)$ is the positive square root of
\begin{equation}
Y_A^\sharp(\lambda)^2 = 
\left[\gamma_2^* u_{AA}\omega_{P}^\sharp\right]^\wedge(-\lambda,\lambda)
\quad\hbox{(no sum on $A$)}.
\label{eq:YAl}
\end{equation}
Furthermore, $Y_A^\cdot(\lambda)$ (resp. $Y_A^\Gamma(\lambda)$) 
decays rapidly as $\lambda\to +\infty$ (resp. $\lambda\to -\infty$) 
and is of polynomially bounded growth as $\lambda\to-\infty$ (resp.
$\lambda\to +\infty$).
\end{Lem}
{\noindent\bf Remarks.} (i) The right-hand side of Eq.~(\ref{eq:YAl}) is nonnegative
because $u_{AA}\omega_P^\sharp$ is of positive type as a scalar
bi-distribution, and this property is inherited under pull-back by
$\gamma_2$ (see Theorem~2.2 in \cite{AGWQI}). In fact, the calculation
\begin{equation}
u_{AB}\omega_P^\sharp(f^A\otimes \overline{f^B}) =
\omega_P^\sharp(f^Au_A\otimes (f^Bu_B)^+)=\omega_0(\Psi(f^Au_A)\Psi(f^Bu_B)^*)
\ge 0
\end{equation}
(summing on $A$ and $B$) for $f^A\in\CoinX{M}$ ($A=1,\ldots,4$) shows that
$u_{AB}\omega_P^\sharp$ is of positive type as a matrix-valued
distribution. Positive type of $u_{AA}\omega_P^\sharp$ follows in consequence. 
A similar argument (using~(\ref{eq:omsnf}) and the property $Q\ge 0$ for
$Q\in\Had(\HH,\Gamma)$) shows that
$u_{AB}\omega^{\cdot\,\Gamma\,\cdot}$,
$u_{AB}\omega^{\Gamma\,\cdot\,\Gamma}$ and their pull-backs by
$\gamma_2$ share the matrix positive type property.
\\[6pt]
{\noindent(ii)} The statements on the growth of $Y_A^\sharp$ are
obtained from the Paley-Wiener-Schwartz theorem~\cite{Hormander1} which entails that
the Fourier transform of a compactly supported distribution is of at
worst polynomial growth. Below, we will frequently use the fact that the
product of a rapidly decaying function and one of polynomial growth is
itself rapidly decaying.

Because the bounds obtained in Lemma~\ref{lem:dineq} exhibit different behaviour
in the four quadrants $C_1,\ldots,C_4$ of the
$(\lambda,\lambda')$-plane it is convenient to decompose the averaged
energy density~(\ref{eq:Idef}) as $\cI=\sum \cI_k$, where $\cI_k$ is the
contribution arising from quadrant $C_k$. We proceed to bound the $\cI_k$ in turn. 

Starting with the second and fourth quadrants $C_2=\RR^-\times\RR^+$ and
$C_4=\RR^+\times\RR^-$, Lemma~\ref{lem:dineq} yields the
$Q$-independent bound
\begin{equation}
\left|W_{AB}^{(\omega_Q)}(\lambda,\lambda')\right|\le \sum_{\sharp\,\flat}
X_{AB}^{\sharp\,\flat}(\lambda,\lambda')
\end{equation}
in which each summand on the right-hand side is of at worst polynomial
growth. This may be combined with 
the rapid decay of $J^{AB}$ away from the diagonal to yield
the following $Q$-independent bound on the contribution from these quadrants:
\begin{equation}
|\cI_2+\cI_4| \le \int_{C_2\cup C_4} d\lambda\,d\lambda' \left|
  J^{AB}(\lambda,\lambda')\right| 
\sum_{\sharp\,\flat} X_{AB}^{\sharp\,\flat}(\lambda,\lambda') <\infty\,.
\end{equation}

We are left with the first and third quadrants. Since $J^{AB}$ exhibits
polynomial growth along the diagonal, the previous argument will not
allow us to bound all the terms arising from the decomposition~(\ref{eq:dec}). To see
this, note that Lemma~\ref{lem:dineq} applied to the $P^\Gamma Q P^\Gamma$ term
$W_{AB}^{\Gamma\,\cdot\,\Gamma}$ 
gives a bound $X_{AB}^{\Gamma\,\Gamma}$ growing polynomially\footnote{Note that it is the
bound which is polynomially growing; we expect that 
$W_{AB}^{\Gamma\,\cdot\,\Gamma}$ is actually decaying.} in all directions in the first
quadrant. Similarly, the bound $X_{AB}^{\cdot\,\cdot}$ for the $PQ^\Gamma P$ term is
polynomially growing in the third quadrant. However,
Lemma~\ref{lem:dineq} suffices to bound the other contributions to
$\cI_1$ and $\cI_3$ because at least one factor in the relevant
$X_{AB}^{\sharp\,\flat}$ is rapidly decaying. Thus 
\begin{equation}
\left| \cI_1 - \cR_1\right| 
\le \int_{C_1} d\lambda\,d\lambda' \left| J^{AB}(\lambda,\lambda')\right|
\left[X_{AB}^{\cdot\,\cdot}(\lambda,\lambda') + 
X_{AB}^{\cdot\,\Gamma}(\lambda,\lambda')
+X_{AB}^{\Gamma\,\cdot}(\lambda,\lambda')
\right] <\infty
\end{equation}
and
\begin{equation}
\left| \cI_3 - \cR_3 \right| 
\le \int_{C_3} d\lambda\,d\lambda' \left| J^{AB}(\lambda,\lambda')\right|
\left[X_{AB}^{\Gamma\,\Gamma}(\lambda,\lambda') + 
X_{AB}^{\cdot\,\Gamma}(\lambda,\lambda')
+X_{AB}^{\Gamma\,\cdot}(\lambda,\lambda')
\right] <\infty\,,
\end{equation}
where the remaining terms are
\begin{equation}
\cR_1 = \int_{C_1} d\lambda\,d\lambda' J^{AB}(\lambda,\lambda')
W_{AB}^{\Gamma\,\cdot\,\Gamma}(\lambda,\lambda')
\label{eq:R1def}
\end{equation}
and
\begin{equation}
\cR_3 = -\int_{C_3} d\lambda\,d\lambda' J^{AB}(\lambda,\lambda')
W_{AB}^{\cdot\,\Gamma\,\cdot}(\lambda,\lambda')
\end{equation}
(the leading minus sign arises because it is $-PQ^\Gamma P$ which
appears in~(\ref{eq:decomp})). The quantities $\cR_1$ and $\cR_3$ are
real, because $J^{AB}$,  
$W_{AB}^{\Gamma\,\cdot\,\Gamma}$ and 
$W_{AB}^{\cdot\,\Gamma\,\cdot}$ are hermitian matrix kernels. 

To complete the proof of the QWEI it is required to show that $\cR_1$ and
$\cR_3$ are bounded from below independently of $Q$. 
We will present the argument for $\cR_1$ in
detail and indicate how the proof is modified for $\cR_3$. Let 
$\chi_\Lambda$ ($\Lambda\in\RR^+$) be a 
family of smooth, real-valued, nonincreasing functions such that 
$\chi_\Lambda$ equals unity on $[0,\Lambda]$ and vanishes on
$[\Lambda+1,\infty)$. It is clear that
\begin{equation}
\cR_1 = \lim_{\Lambda\to\infty} 
\int_{C_1} d\lambda\,d\lambda'
\chi_\Lambda(\lambda)\chi_\Lambda(\lambda')
J^{AB}(\lambda,\lambda')
W_{AB}^{\Gamma\,\cdot\,\Gamma}(\lambda,\lambda')
\label{eq:R1lim}
\end{equation}
and the cut-off $\chi_\Lambda$ now allows us to interpret the integral
as a trace in the following
way. Define $\sigma:\RR^+\to\RR$ by
\begin{equation}
\sigma(\lambda) =
(1+\lambda^2)^{1/2}[1+|Y^\Gamma(\lambda)|_{\CC^4}]\,,
\end{equation} 
where $|\cdot|_{\CC^4}$ is the usual vector norm on $\CC^4$. Then
$\sigma$ is smooth, positive, bounded away from zero and of polynomially bounded
growth. Next, define operators $\widetilde{J}_\Lambda$ and $\widetilde{W}$ on 
$L^2(\RR^+,d\lambda)\otimes\CC^4$ by
\begin{equation}
(\widetilde{J}_\Lambda\varphi)^A(\lambda) = \int_0^\infty d\lambda'\,
\sigma_\Lambda(\lambda) \sigma_\Lambda(\lambda')
J^{AC}(\lambda,\lambda')\delta_{CB}\varphi^B(\lambda')
\end{equation}
and 
\begin{equation}
(\widetilde{W}\varphi)^B(\lambda') = \delta^{BD}
\int_0^\infty d\lambda''\,
\frac{W_{CD}^{\Gamma\,\cdot\,\Gamma}(\lambda'',\lambda')}{\sigma(\lambda')\sigma(\lambda'')}
\varphi^C(\lambda'')\,,
\end{equation}
where have written $\sigma_\Lambda(\lambda)=\chi_\Lambda(\lambda)
\sigma(\lambda)$. Now $\widetilde{J}_\Lambda$ is Hilbert-Schmidt (due to
the cut-off) and self-adjoint while the properties of $\widetilde{W}$
are summarised in the following proposition, which is proved at the end
of this section.
\begin{Prop} \label{prop:W}
For all $Q\in\Had(\HH,\Gamma)$, $\widetilde{W}$ is a positive trace-class operator with 
\begin{equation}
0\le \Tr \widetilde{W}  \le \frac{\pi}{2}\,.
\label{eq:trbd}
\end{equation}
\end{Prop}
We may therefore rewrite 
Eq.~(\ref{eq:R1lim}) in the form
\begin{equation}
\cR_1 = \lim_{\Lambda\to\infty} \Tr \widetilde{J}_\Lambda \widetilde{W}\,.
\label{eq:R1}
\end{equation}
Introducing an orthornormal basis of eigenvectors $\upsilon_n$ for $\widetilde{W}$,
and using $\widetilde{W}\ge 0$ and~(\ref{eq:trbd}), we have
\begin{eqnarray}
\Tr \widetilde{J}_\Lambda\widetilde{W} &=& \sum_n
\ip{\upsilon_n}{\widetilde{J}_\Lambda\widetilde{W}\upsilon_n} = 
\sum_n
\ip{\upsilon_n}{\widetilde{J}_\Lambda\upsilon_n}\ip{\upsilon_n}{\widetilde{W}\upsilon_n}
\nonumber\\
&\ge& \inf \spec(\widetilde{J}_\Lambda) \sum_n \ip{\upsilon_n}{\widetilde{W}\upsilon_n}
\nonumber\\
&\ge& \inf \spec(\widetilde{J}_\Lambda)\, \Tr \widetilde{W}\nonumber\\
&\ge& \frac{\pi}{2}\min\{0,\inf \spec(\widetilde{J}_\Lambda)\}\,.
\end{eqnarray}
Noting that the right-hand side has no $Q$-dependence, the required
lower bound on $\cR_1$ now follows from the following
proposition, which is proved in Sec.~\ref{sect:J}. 
\begin{Prop} \label{prop:J}
The spectrum of $\widetilde{J}_\Lambda$ is bounded from below uniformly in
$\Lambda$ (with a finite lower bound). 
\end{Prop}

Turning to the integral $\cR_3$, we follow exactly the same argument but
with the single difference that the kernel $J^{AB}$ on
$\RR^-\times\RR^-$ defines an operator on $L^2(\RR^-)\otimes \CC^4$ which may be bounded
{\em above} by a nonnegative quantity; this is compensated by the
leading minus sign in the definition of $\cR_3$. Thus $\cR_1+\cR_3$ has a
finite $Q$-independent lower bound as required and the proof of
Theorem~\ref{thm:main} is complete. $\square$

{\noindent\em Proof of Proposition~\ref{prop:W}:} 
Using Lemma~\ref{lem:dineq} and Cauchy-Schwarz, we first
estimate
\begin{equation}
|(\widetilde{W}\varphi)(\lambda')|_{\CC^4}\le 
\frac{|Y^\Gamma(\lambda')|_{\CC^4}}{(1+{\lambda'}^2)^{1/2}[1+|Y^\Gamma(\lambda')|_{\CC^4}]} \int
d\lambda''\,
\frac{Y_C^\Gamma(\lambda'')}{(1+{\lambda''}^2)^{1/2}[1+|Y^\Gamma(\lambda'')|_{\CC^4}]}
|\varphi^C(\lambda'')| 
\end{equation}
for $\varphi\in L^2(\RR^+)\otimes
\CC^4$, from which we obtain
$\|\widetilde{W}\varphi\|\le \frac{1}{2}\pi\|\varphi\|$. Thus
$\widetilde{W}$ is bounded. A short calculation shows that
\begin{equation}
\ip{\varphi}{\widetilde{W}\varphi} = \left(\gamma_2^*
u_{CB}\omega^{\Gamma\,\cdot\,\Gamma}\right)
(\psi^C\otimes \overline{\psi^B})\ge 0
\end{equation}
for $\varphi\in\CoinX{\RR^+}\otimes\CC^4$ where 
\begin{equation}
\psi(\tau) =\frac{1}{2\pi}\left[ \frac{1}{\sigma}\varphi\right]^\vee(\tau)
\in \SSS(\RR)\otimes\CC^4\,,
\end{equation}
and we have used the matrix
positive type property and compact support of 
$\gamma_2^*u_{CB}\omega^{\Gamma\,\cdot\,\Gamma}$. Thus $\widetilde{W}$
is positive. In order to show that $\widetilde{W}$ is trace-class it is
now enough (by the lemma\footnote{For this
purpose, we regard $L^2(\RR^+,d\lambda)\otimes\CC^4$ as
$L^2(X,d\lambda\otimes d\mu)$ where $X$ is the locally compact space 
$\RR^+\times\ZZ_4$ and $\mu$ is the counting measure on $\ZZ_4$. The
measure $d\lambda\otimes d\mu$ is a Baire measure and the lemma may be
applied.} following Theorem X1.31 in \cite{RS3}) to show that
the formal trace of $\widetilde{W}$ is finite (whereupon the formal
trace is indeed the trace). But this is just
\begin{equation}
\int_0^\infty d\lambda\, \sigma(\lambda)^{-2}\delta^{BC}W_{BC}(\lambda,\lambda)
\le \int_0^\infty d\lambda\, 
\frac{|Y^\Gamma(\lambda)|_{\CC^4}^2}{(1+\lambda^2)(1+|Y^\Gamma(\lambda)|_{\CC^4})^2}
\le \frac{\pi}{2}\,,
\end{equation}
where we have used Lemma~\ref{lem:dineq} and the fact that the diagonal of
the kernel of a positive trace-class operator 
is positive. Thus $\widetilde{W}$ is trace-class on
$L^2(\RR^+)\otimes\CC^4$; putting this together with the positivity property,
we have $0\le \Tr\widetilde{W}\le \pi/2$ as required. $\square$

\section{Proof of Lemma~\ref{lem:dineq}} \label{sect:pfdineq}

We first establish the existence of the quantities
$W_{AB}^{\sharp\,\natural\,\flat}$ and $Y_A^\sharp(\lambda)$ defined
by~(\ref{eq:WABsnf}) and~(\ref{eq:YAl}).
Using self-adjointness of $P^\sharp$, Cauchy-Schwarz and 
$\|Q\|\le 1$ (following from
property (I) of Sec.~\ref{sect:Hadstate} for $Q\in\Had(\HH,\Gamma)$)
we obtain from~(\ref{eq:omsnf}) the inequality 
\begin{equation}
| \omega^{\sharp\,\natural\,\flat}(f_1\otimes f_2^+) |^2  
\le \left\|P^\sharp \Gamma\left[\begin{array}{c} 0\\
f_1\end{array}\right]
\right\|^2\, \left\|P^\flat \left[\begin{array}{c} f_2^+\\
0\end{array}\right]\right\|^2
= \omega_{P}^\sharp(f_1\otimes f_1^+)\omega_{P}^\flat(f_2\otimes f_2^+)
\label{eq:topineq}
\end{equation}
for any $f_i\in\DDsp$ ($i=1,2$). This inequality underlies the following 
lemma, which will be proved at the end of this section.
\begin{Lem} \label{lem:WF}
The wave-front set of $\omega^{\sharp\,\natural\,\flat}$ satisfies 
\begin{equation}
\WF(\omega^{\sharp\,\natural\,\flat})\subset 
\bigcup_{p,p'\in M}
\{p\}\times\Nn^\sharp_{p} \times \{p'\} \times -\Nn^\flat_{p'}
\label{eq:topwf}
\end{equation}
as a subset of $T^*(M\times M)$. 
\end{Lem}
This is by no means a sharp estimate of
the wave-front set, but it will suffice for our purposes. 

Defining $u_{AB}$ as in~(\ref{eq:uAB}),
$u_{AB}\omega^{\sharp\,\natural\,\flat}$ is a scalar bi-distribution
with wave-front set contained in the right-hand-side of
Eq.~(\ref{eq:topwf}). Now by Theorem~2.5.11${}'$
in~\cite{Hfio1}, the pull-back 
$\gamma_2^*u_{AB}\omega^{\sharp\,\natural\,\flat}$ exists provided the
intersection of its wave-front set with the set of normals
$N_{\gamma_2}$ of $\gamma_2$ is empty. One may show that 
\begin{equation}
N_{\gamma_2} = \{(\gamma(\tau),\xi;\gamma(\tau'),\xi')\in T^*(M\times M)\mid \xi_a
u^a(\tau)=\xi'_{b'}u^{b'}(\tau')=0\}
\end{equation}
(see Sec.~3 of~\cite{AGWQI} where a corresponding argument is given).
This has trivial intersection with
$\WF(u_{AB}\omega^{\sharp\,\natural\,\flat})$, because no null covector
can annihilate a timelike vector. Accordingly, 
$\gamma_2^*u_{AB}\omega^{\sharp\,\natural\,\flat}$ exists in $\DD'(\RR^2)$, 
and (again by Theorem~2.5.11${}'$ in~\cite{Hfio1}) its 
wave-front set obeys
\begin{equation}
\WF(\gamma_2^*u_{AB}\omega^{\sharp\,\natural\,\flat})\subset \RR
\times \RR^\sharp \times  
\RR \times -\RR^\flat  \,.
\label{eq:pbwf}
\end{equation}
Since $u_{AB}$ is compactly supported, we may take Fourier transforms
and conclude that the $W_{AB}^{\sharp\,\natural\,\flat}$ do indeed exist. 
Exactly the same argument, using the microlocal spectrum
condition~(\ref{eq:uSC2}) in place of~(\ref{eq:topwf}), shows that
the pull-backs $\gamma_2^*u_{AB}\omega_P^{\sharp}$ exist with
\begin{equation}
\WF(\gamma_2^*u_{AB}\omega_P^{\sharp})\subset \RR
\times \RR^\sharp \times  
\RR \times -\RR^\sharp  \,.
\label{eq:pbwf2}
\end{equation}

It remains to prove Lemmas~\ref{lem:dineq} and~\ref{lem:WF}.

{\noindent\em Proof of Lemma~\ref{lem:dineq}:} An argument using
regularising sequences in analogy with the proof of Theorem~2.2
in~\cite{AGWQI} shows that the
inequality~(\ref{eq:topineq}) is inherited by the pull-back and becomes
\begin{equation}
|\gamma_2^* u_{AB}\omega^{\sharp\,\natural\,\flat}(f_1\otimes \overline{f_2}) |^2 \le
\gamma_2^*u_{AA}\omega_{P}^\sharp(f_1\otimes \overline{f_1})\,
\gamma_2^*u_{BB}\omega_{P}^\flat(f_2\otimes \overline{f_2})\ \ \ \forall\
f_1,f_2\in\CoinX{\RR}
\end{equation}
(no sum on either $A$ or $B$). Substituting $f_1(t)= {\rm e}^{-it\lambda}$,
$f_2(t')={\rm e}^{-it'\lambda'}$, the required bounds~(\ref{eq:dineq})
are obtained.  

As $\gamma_2^* u_{AA}\omega^\cdot_{P}$ is compactly supported, its set of
singular directions (those directions in which its
Fourier transform fails to decay rapidly) is given by
\begin{eqnarray}
\Sigma(\gamma_2^* u_{AA}\omega^\cdot_{P}) &=& \{
(\lambda,\lambda')\mid (\tau,\lambda;\tau',\lambda')\in
\WF(\gamma_2^* u_{AA}\omega^\cdot_{P})~\hbox{for
some}~(\tau,\tau')\in\RR^2\} \nonumber\\
&=& \RR^+\times\RR^-
\end{eqnarray}
(see Proposition~8.1.3 in~\cite{Hormander1}). Thus $(-1,1)$ is not a
singular direction for $\gamma_2^* u_{AA}\omega^\cdot_{P}$ and we
deduce that $Y_{A}^\cdot(\lambda)$
is rapidly decaying at $\lambda\to+\infty$ and of polynomially bounded
growth as $\lambda\to-\infty$ by the Paley-Wiener-Schwartz
theorem~\cite{Hormander1}. An analogous argument shows that
$Y^\Gamma_{A}(\lambda)$ decays rapidly as $\lambda\to-\infty$ and is
polynomially bounded as $\lambda\to +\infty$. $\square$ 

{\noindent\em Proof of Lemma~\ref{lem:WF}:} Suppose
$(p_1,\xi_1;p_2,-\xi_2)\in \WF(\omega^{\sharp\,\natural\,\flat})$. We will
show that
\begin{equation}
(p_1,\xi_1;p_1,-\xi_1)\in \WF(\omega_P^\sharp) \qquad{\rm and}\qquad
(p_2,\xi_2;p_2,-\xi_2)\in \WF(\omega_P^\flat)\,,
\label{eq:goal}
\end{equation}
from which the required result now follows by the microlocal spectrum
condition~(\ref{eq:uSC2}). To prove~(\ref{eq:goal}) fix charts $(U_i,\kappa_i)$ with
$p_i\in U_i$ ($i=1,2$), and define $k_i$ so that $\xi_i =
{}^{t}\kappa_i'(p_i)k_i$. Let $\chi_i$ be arbitrary smooth spinor fields
compactly supported in $U_i$ with $\chi_i(p_i)\not =0$. 
By definition of the wave-front set on manifolds, $(k_1,-k_2)$
is a singular direction for $\chi_{12}\omega^{\sharp\,\natural\,\flat}
\circ \kappa_{12}^{-1}$, where we use the notation
\begin{equation}
\chi_{ij} = \chi_i\otimes \chi_j^+;\qquad
\kappa_{ij}=\kappa_i\times\kappa_j\,.
\end{equation}
Thus, if $V_j$ are arbitrary conical neighbourhoods of $k_j$, 
there must exist $k'_j\in V_j$ such that
\begin{equation}
\left[
\chi_{12}\omega^{\sharp\,\natural\,\flat}\circ\kappa_{12}^{-1}\right]^\wedge
(\alpha k'_1,-\alpha k'_2)
\end{equation}
is not of rapid decay as $\alpha\to+\infty$. Applying~(\ref{eq:topineq}) to 
\begin{equation}
f_j = \left(\frac{1}{\sqrt{-|\gb|}\circ\kappa_j^{-1}}\chi_j
{\rm e}^{i(\,.\,)\alpha k'_j}\right)\circ \kappa_j\,,
\end{equation}
and recalling~(\ref{eq:distdef}), we obtain 
\begin{equation}
\left|\left[
\chi_{12}\omega^{\sharp\,\natural\,\flat}\circ\kappa_{12}^{-1}\right]^\wedge
(\alpha k'_1,-\alpha k'_2)\right|
\le\left[
\chi_{11}\omega_P^{\sharp}\circ\kappa_{11}^{-1}\right]^\wedge
(\alpha k'_1,-\alpha k'_1) \,
\left[
\chi_{22}\omega_P^{\flat}\circ\kappa_{22}^{-1}\right]^\wedge
(\alpha k'_2,-\alpha k'_2)
\end{equation}
and since the left-hand side is not of rapid decrease as
$\alpha\to\infty$ we conclude that
the same must be true of both factors on the right-hand side. 
Hence (as the
$V_j$ were arbitrary) $(k_1,-k_1)$ is a singular direction for
$\chi_{11}\omega_P^\sharp\circ\kappa_{11}^{-1}$ and
$(k_2,-k_2)$ is a singular direction for
$\chi_{22}\omega_P^\flat\circ\kappa_{22}^{-1}$. Letting the support of each 
$\chi_i$ shrink to $\{p_i\}$, we have
\begin{equation}
(\kappa(p_1),k_1;\kappa(p_1),-k_1)\in\WF(\omega_P^\sharp\circ\kappa_{11}^{-1})
\end{equation}
and
\begin{equation}
(\kappa(p_2),k_2;\kappa(p_2),-k_2)\in\WF(\omega_P^\flat\circ\kappa_{22}^{-1})\,,
\end{equation}
from which~(\ref{eq:goal}) follows immediately. $\square$

\section{Proof of Proposition~\ref{prop:J}}\label{sect:J}

We now prove that the operators $\widetilde{J}_\Lambda$ on
$L^2(\RR^+,d\lambda)\otimes\CC^4$ 
are bounded from below uniformly in $\Lambda$. To begin, we consider the
related operator $K_\Lambda$ acting on $L^2(\RR^+,d\lambda)$ by
\begin{equation}
(K_\Lambda\varphi)(\lambda) = \int_0^\infty d\lambda
\frac{\sigma_\Lambda(\lambda)\sigma_\Lambda(\lambda')}{8\pi^2}
(\lambda+\lambda') \widehat{f}(\lambda-\lambda') \varphi(\lambda')\,.
\end{equation}
If the spin-connection terms vanished, $\widetilde{J}_\Lambda$ would be
equal to $K_\Lambda\otimes\II$. Our analysis of $K_\Lambda$ is based
on the following identity.
\begin{Lem} If $f=g^2$ for real-valued $g\in\CoinX{\RR}$, then
\begin{equation}
(\lambda+\lambda')\widehat{f}(\lambda-\lambda') = 
\int_{-\infty}^\infty\frac{d\mu}{\pi}
\mu \widehat{g}(\lambda-\mu)\overline{\widehat{g}(\lambda'-\mu)}\,.
\label{eq:Kid}
\end{equation}
\end{Lem}
{\noindent\em Proof:} Note first that the right-hand side (RHS) exists
for each $\lambda,\lambda'\in\RR$. Changing variables to
$\nu=\mu-(\lambda+\lambda')/2$ and writing $\zeta = (\lambda-\lambda')/2$
\begin{eqnarray}
\hbox{RHS of~(\ref{eq:Kid})} &=& \int_{-\infty}^\infty \frac{d\nu}{\pi}\,
\left(\frac{\lambda+\lambda'}{2}+\nu\right)
\widehat{g}(\zeta-\nu)\widehat{g}(\zeta+\nu)
\nonumber\\
&=& (\lambda+\lambda')(\widehat{g}\star\widehat{g})(2\zeta) \nonumber\\
&=& (\lambda+\lambda') \widehat{f}(\lambda-\lambda')
\end{eqnarray}
as required, where we have also used
$\overline{\widehat{g}(u)}=\widehat{g}(-u)$, the fact that $\nu 
\widehat{g}(\zeta-\nu)\widehat{g}(\zeta+\nu)$ is odd, and a further
change of variables. In addition, we have used $\star$ to denote the 
convolution $(h_1\star h_2)(\lambda) = \int d\lambda'/(2\pi)
h_1(\lambda-\lambda')h_2(\lambda')$. 
$\square$

It follows from this identity that the kernel of $K_\Lambda$ may be rewritten
in the form
\begin{equation}
K_\Lambda(\lambda,\lambda') =  
\frac{\sigma_\Lambda(\lambda)\sigma_\Lambda(\lambda')}{8\pi^3}
\int_{-\infty}^\infty d\mu\,
\mu \widehat{g}(\lambda-\mu)\overline{\widehat{g}(\lambda'-\mu)}\,.
\end{equation}
We now define $K_\Lambda^\pm$ to have the kernels
\begin{equation}
K^+_\Lambda(\lambda,\lambda') = 
\frac{\sigma_\Lambda(\lambda)\sigma_\Lambda(\lambda')}{8\pi^3}
\int_{-\infty}^\infty d\mu\,
|\mu| \widehat{g}(\lambda-\mu)\overline{\widehat{g}(\lambda'-\mu)}
\end{equation}
and
\begin{equation}
K^-_\Lambda(\lambda,\lambda') =  -
\frac{2\sigma_\Lambda(\lambda)\sigma_\Lambda(\lambda')}{8\pi^3}
\int_{-\infty}^0 d\mu\,
\mu \widehat{g}(\lambda-\mu)\overline{\widehat{g}(\lambda'-\mu)}\,.
\end{equation}
The integrals in these kernels are bounded on compact subsets of
$\RR^+\times\RR^+$, so the cut-off functions $\sigma_\Lambda$ ensure
that $K_\Lambda$ and $K_\Lambda^\pm$ are Hilbert-Schmidt. Clearly
$K_\Lambda=K^+_\Lambda-K_\Lambda^-$; furthermore, the easily proven identity
\begin{equation}
\ip{\varphi}{K_\Lambda^-\varphi} = 
\int_0^{\infty} d\mu\, \frac{\mu}{4\pi^3}\left|
\int_0^\infty
d\lambda'\,\overline{\widehat{g}(\lambda'+\mu)}\sigma_\Lambda(\lambda')
\varphi(\lambda')\right|^2
\label{eq:Kcalc}
\end{equation}
(valid, say, for $\varphi\in\CoinX{\RR^+}$) shows that $K_\Lambda^-$ is
positive. A similar argument establishes positivity of $K^+_\Lambda$.

Our aim is now to find a bound on $K_\Lambda^-$ which will allow a bound
uniform in $\Lambda$ to be obtained. (The operator $K_\Lambda^+$ is
bounded for each $\Lambda$, e.g., by its Hilbert-Schmidt norm, but
becomes unbounded in the limit $\Lambda\to\infty$). 
Regarding the inner integral in~(\ref{eq:Kcalc})
as an $L^2$-inner product and applying Cauchy-Schwarz, we obtain
\begin{equation}
\ip{\varphi}{K_\Lambda^-\varphi} \le C_\Lambda
\|\varphi\|^2\,,
\label{eq:KLminusbd}
\end{equation}
where
\begin{eqnarray}
C_\Lambda &=& \frac{1}{4\pi^3}
\int_{\RR^+\times\RR^+}d\mu\,d\lambda'\,
\mu\left|\overline{\widehat{g}(\mu+\lambda')} \sigma_\Lambda(\lambda')\right|^2
\nonumber\\
&=& \int_0^\infty du |\widehat{g}(u)|^2 F_\Lambda(u)
\end{eqnarray}
and
\begin{equation}
F_\Lambda(u) = \frac{1}{4\pi^3}\int_0^u d\lambda' (u-\lambda')\sigma_\Lambda(\lambda')^2
\label{eq:FL}
\end{equation}
is bounded and nonnegative. 
Let us observe that this step depends in an essential way on the
fact that, for $\mu>0$, the argument of $\widehat{g}$
in~(\ref{eq:Kcalc}) is bounded away from zero, together with the rapid
decay property of $\widehat{g}$.
 
The above analysis entails that $-C_\Lambda$ is a lower bound for
$K_\Lambda$, but this is certainly not the sharpest bound. In fact
essentially the same argument applies if $K_\Lambda^-$ is replaced by 
$\frac{1}{2}K_\Lambda^-$ and $K^+_\Lambda$ is adjusted to maintain 
$K_\Lambda=K^+_\Lambda-K_\Lambda^-$, with the conclusion that
$-C_\Lambda/2$ is also a lower bound for $K_\Lambda$. The convenience of
the choices made above is that, as we now show, the operator 
$L_\Lambda=\widetilde{J}_\Lambda-K_\Lambda\otimes\II$ is form bounded
relative to $K_\Lambda^+$ with relative bound no greater than
$\frac{1}{2}$. To this end, we first use the convolution theorem to write
\begin{equation}
\ip{\varphi}{L_\Lambda\varphi} = \frac{1}{2}\int_{-\infty}^\infty d\tau
[\sigma_\Lambda\varphi]^\vee(\tau)^\dagger f(\tau)\theta(\tau) [\sigma_\Lambda\varphi]^\vee(\tau)
\end{equation}
for $\varphi\in\CoinX{\RR^+}\otimes\CC^4$,
where $\dagger$ denotes the matrix hermitian conjugate. Setting
$\displaystyle C'= \sup_{\tau\in\RR} \|\theta(\tau)\|_{\CC^4}$, we then estimate
\begin{eqnarray}
|\ip{\varphi}{L_\Lambda\varphi}| &\le &  
\frac{C'}{2} \int_{-\infty}^\infty d\tau\, 
\left| (g[\sigma_\Lambda\varphi]^\vee)(\tau)
\right|_{\CC^4}^2\nonumber\\
&\le& \frac{C'}{2}\int_{-\infty}^\infty
\frac{d\mu}{2\pi} \left| (g[\sigma_\Lambda\varphi]^\vee)^\wedge(\mu) 
\right|_{\CC^4}^2 \nonumber \\
&=& \frac{C'}{2} \int_{-\infty}^\infty
\frac{d\mu}{2\pi} \left|\int_0^\infty\frac{d\lambda'}{2\pi}
\widehat{g}(\mu-\lambda')\sigma_\Lambda(\lambda')\varphi(\lambda')
\right|_{\CC^4}^2\,.
\end{eqnarray}
By comparison with the definition of $K_\Lambda^+$, this implies
\begin{equation}
|\ip{\varphi}{L_\Lambda\varphi}| \le \frac{1}{2}
\ip{\varphi}{(K_\Lambda^+\otimes\II)\varphi} + 
\frac{C'}{16\pi^3} \int_{|\mu|<C'}
d\mu\, \left|\int_0^\infty d\lambda'
\widehat{g}(\mu-\lambda')\sigma_\Lambda(\lambda')\varphi(\lambda')
\right|_{\CC^4}^2\,.
\end{equation}
Using a similar argument to that used to obtain~(\ref{eq:KLminusbd}), the last term may be bounded
by $C''_{\Lambda}\|\varphi\|^2$ where
\begin{eqnarray}
C''_\Lambda &=&\frac{C'}{16\pi^3}\int_{|\mu|<C'} d\mu\,
\int_0^\infty d\lambda' |\overline{\widehat{g}(\mu+\lambda')}
\sigma_\Lambda(\lambda')|^2\nonumber\\
&=& \int_{-C'}^\infty du\, |\widehat{g}(u)|^2 G_\Lambda(u)
\end{eqnarray}
and the bounded nonnegative function $G_\Lambda$ is given on $[-C',\infty)$ by
\begin{equation}
G_\Lambda(u) = \frac{C'}{16\pi^3} \int_{\max\{u-C',0\}}^{u+C'}
d\lambda' \sigma_\Lambda(\lambda')^2\,.
\label{eq:GL}
\end{equation}
Thus $L_\Lambda$ is form bounded relative to $K_\Lambda^+$ as claimed
above. Since $\widetilde{J}_\Lambda=K_\Lambda^+-K_\Lambda^-+L_\Lambda$, we have
\begin{eqnarray}
|\ip{\varphi}{(\widetilde{J}_\Lambda-K_\Lambda^+\otimes\II)\varphi}|
&\le & |\ip{\varphi}{L_\Lambda\varphi}| + |\ip{\varphi}{K^-_\Lambda\varphi}|
\nonumber\\
&\le &
\frac{1}{2} 
\ip{\varphi}{(K_\Lambda^+\otimes\II)\varphi}
+(C_\Lambda+C''_{\Lambda})\|\varphi\|^2
\end{eqnarray}
and, as $K_\Lambda^+$ is positive, 
it follows that $\widetilde{J}_\Lambda$ is bounded below by
$-(C_\Lambda+C''_\Lambda)$. 

To complete the proof of Proposition~\ref{prop:J} we must show that 
$C_\Lambda$ and $C_\Lambda''$ may both
be bounded above uniformly in $\Lambda$. This holds because
$\sigma_\Lambda(\lambda)$ is pointwise dominated by $\sigma(\lambda)$
and thus $F_\Lambda$ and $G_\Lambda$ are pointwise dominated by the
functions $F$ and $G$ obtained by replacing $\sigma_\Lambda$ by $\sigma$
in~(\ref{eq:FL}) and~(\ref{eq:GL}). Furthermore, $F$ and $G$ are of
polynomially bounded growth so we obtain the bounds
\begin{equation}
C_\Lambda\le \int_0^\infty du |\widehat{g}(u)|^2 F(u) <\infty
\end{equation}
and
\begin{equation}
C_\Lambda''\le \int_{-C'}^\infty du\, |\widehat{g}(u)|^2 G(u)<\infty
\end{equation}
for all $\Lambda$, where the rapid decay property of $\widehat{g}$ has
been used. 

\section{Majorana fields} \label{sect:Maj}

In this section we will indicate the changes required in Sec.~\ref{sect:psed}
when treating Majorana fields. Suppose that $\gamma_0,\ldots,\gamma_3$
belong to a Majorana representation and that $\HH$ is the completion
of $\DDco/{\rm ker}\,S_\lhd$ with scalar product given by $(h,h')_S =
(S_\lhd h,h')$, cf.\ Remark (iii) at the end of Sec.~\ref{sect:DirEq}. The field
operators are then given by 
\begin{equation}
 \Psi(h) = B(q(h));\qquad \Psi^+(f)=\Psi(f^+)^*=\Psi(\Gamma f^+)
\qquad h \in \DDco,~f\in\DDsp
\label{eq:Majops}
\end{equation}
where $q: \DDco \to \DDco/{\rm ker}\,S_\lhd$ is the quotient map. Now,
as in Sec.~\ref{sect:psed} one may choose, in a tubular neighbourhood of any
timelike curve $\gamma$, induced frames $(e_0,\ldots,e_3)$ and $(E_A)$
so that $e_0|_{\gamma} = u$, the tangent of $\gamma$ in proper time
parametrization. However, it is now
convenient to set $v_A = i E_A$, for then we have the two properties
\begin{equation}
\delta^{AB} v_A \otimes v_B^+ = \gamma_0 \qquad{\rm and}\qquad \Gamma v_A^+ = v_A^+
\end{equation}
as we are working in a Majorana representation. 

To quantize the point-split energy density \eqref{iii4},
we may substitute from~(\ref{eq:Majops}) and use the formula
$\Gamma[\nabla\cdot (e_0 v_A f)]^+ = \nabla\cdot (fe_0 v_A^+)$
for $f\in\DD(M)$ to obtain 
\begin{equation}
T(f\otimes g) = \delta^{AB}\frac{i}{2}\left(
\Psi(\nabla\cdot [fe_0 v_A^+] ) \Psi(v_B^+ g) 
-\Psi(f v_A^+) \Psi(\nabla\cdot[e_0 v_B^+ g])
\right)
\end{equation}
as the replacement for \eqref{iii6}. The CAR's may be used to show that
$\langle:T:\rangle_\omega(f\otimes g)$ is symmetric
in $f,g$. 

Using the formula $e_0\cdot\nabla v_B^+ = \overline{{\sigma_{0}}^C_B}
v_C^+$, it follows that $\langle T \rangle_\omega$ may be expressed in
the form
\begin{equation}
\langle T \rangle_\omega =
\LL^{AB} v_A^+\otimes v_B^+\omega_{2}\,,
\end{equation}
where $\omega_2(h,h')=\omega(\Psi(h)\Psi(h'))$ is the two-point function
and 
\begin{equation}
\LL^{AB}= \frac{1}{2}
\left(\II\otimes ie_0\cdot\nabla-ie_0\cdot\nabla\otimes\II \right) \delta^{AB}
+ \frac{1}{2}\Theta^{AB} \,,
\end{equation}
with
\begin{equation}
\Theta^{AB} =
i [\delta^{CB}\overline{\sigma_{0}{}^{A}{}_C}\otimes\II -\II\otimes
\delta^{AC}\overline{\sigma_{0}{}^B{}_C}]\,.
\end{equation}
The components of
$\sigma_0$ are real in a Majorana representation 
and it follows that $\theta^{AB}(\tau)=\gamma_2^*\Theta^{AB}(\tau,\tau)$
is hermitian for each $\tau$. Moreover, $v_A^+\otimes v_B^+ \omega_2$ is
of matrix positive type as shown by the calculation
\begin{equation}
(v_A^+\otimes v_B^+\omega_2)(\overline{f^A}\otimes f^B) =
\omega_2(\Gamma [v_A^+ f^A]\otimes v_B^+ f^B)=
\omega (\Psi(v_A^+ f)^*\Psi(v_B^+ f^B))\ge 0
\end{equation}
in which we have used $\Gamma v_A^+=v_A^+$ and~(\ref{eq:Majops}). 

{}From this point onwards, one may proceed to define $\langle :\rho :
\rangle_{\omega}$ as in Sec.\ 3, and the statement and proof of Thm.\
4.1 carry over (apart from some obvious changes) upon observing that Hadamard
states $\omega$ of the Majorana field obey
\begin{equation}
\WF(\omega_2) = \left\{(p,\xi;p',-\xi')\in\dot{T}^*(M\times M) \mid (p,\xi)\sim
(p',\xi');\quad \xi\in \Nn_{p}^+ \right\}
\end{equation}
(cf.\ \cite{SahlmVer2}, note again that in this
reference a different sign convention for the Fourier transform was
chosen which results in the opposite form of the wave-front set there).
\section{Conclusion}

In this paper, we have established general QWEIs for the Dirac and
Majorana fields in
globally hyperbolic spacetimes. We conclude with various remarks. 
First, these QWEIs hold despite the fact that the `classical' Dirac
equation fails to obey the weak energy condition. This is encouraging
evidence that QWEIs are a widespread feature of all quantum field
theories and that they are the correct replacement for the classical
energy conditions. It would be interesting to understand whether QWEIs 
can be obtained in a general axiomatic setting. 

A second point was posed to us by Buchholz (private communication): 
given a weight in $\WW$ the corresponding averaged energy density 
may be defined as a symmetric operator on a suitable dense domain (such
as the domain of microlocal smoothness introduced in~\cite{BF}) in
some Hilbert space representation. The force of our result (and the
corresponding result in~\cite{AGWQI}) is that
such operators are semibounded, and therefore admit self-adjoint
extensions (in particular, the Friedrichs extension). Can one give any
interpretation to the evolution generated by this operator? The answer
is not clear, but we speculate that there could be a connection with the
dynamics discussed by Keyl~\cite{Keyl} in his recent study of quantum fields on
timelike curves. The connection is somewhat tentative (in particular,
the role of the weight must be understood), and would only
be expected to hold under restricted conditions such as for static trajectories
in static spacetimes. Nonetheless, it remains an intriguing possibility. 

Finally, although our approach does lead to explicit lower bounds on the
various contributions to the 
averaged energy density, these are not expected to be optimal. We
hope to return to this question elsewhere. 

{\noindent\em Acknowledgments:} We thank the organisers of the meeting
on Microlocal Analysis and Quantum Field Theory at the Mathematisches
Forschungsinstitut Oberwolfach, where
this work was commenced. CJF thanks Simon Eveson, Alfredo Calvo Pereira
and Stefan Hollands for useful discussions, and is grateful to the
Institut f\"ur Theoretische Physik in G\"ottingen for hospitality in the
later stages of the work. We also thank Detlev Buchholz for raising 
the issue discussed above.  
The work of CJF was assisted by a grant from the Nuffield Foundation.


\end{document}